\documentclass[aps,prl,twocolumn,reprint,floatfix, superscriptaddress]{revtex4-2}

\usepackage[normalem]{ulem}
\usepackage[utf8]{inputenc}
\usepackage{graphicx,amsmath,amssymb,braket, dsfont}
\graphicspath{{figures/}}
\usepackage{color}
\definecolor{darkgreen}{rgb}{0.0, 0.4, 0.26}

\usepackage{physics}

\definecolor{myblue}{rgb}{0.2,0.2,0.8}
\definecolor{myzard}{cmyk}{0,0,0.05,0}
\definecolor{mywhite}{rgb}{1,1,1}
\definecolor{mywhite}{rgb}{1,1,1}
\definecolor{myred}{rgb}{1,0.,0.3}
\usepackage[colorlinks=true,citecolor=myblue,linkcolor=myred]{hyperref}

\def\be{\begin{equation}}
	\def\ee{\end{equation}}
\def\ba{\begin{align}}
	\def\enda{\end{align}}
\def\bi{\begin{itemize}}
	\def\ei{\end{itemize}}

\def\rr{\mathbf{r}}

\def\kk{\mathbf{k}}

\def\aa{\mathbf{a}}

\begin{document}
	
	\title{Tunable Directional ERmission and Collective Dissipation with Quantum Metasurfaces}
	
	\author{D.~Fern\'andez-Fern\'andez}
	\affiliation{Institute of Fundamental Physics IFF-CSIC, Calle Serrano 113b, 28006 Madrid, Spain.}
	\affiliation{Instituto de Ciencia de Materiales de Madrid ICMM-CSIC, 28049 Madrid, Spain.}
	
	\author{A.~Gonz\'alez-Tudela}
	\email{a.gonzalez.tudela@csic.es}
	\affiliation{Institute of Fundamental Physics IFF-CSIC, Calle Serrano 113b, 28006 Madrid, Spain.}
	
	\begin{abstract}
		Subwavelength atomic arrays, recently labeled as quantum metamaterials, have emerged as an exciting platform for obtaining novel quantum optical phenomena. The strong interference effects in these systems generate subradiant excitations that propagate through the atomic array with very long lifetimes. Here, we demonstrate that one can harness these excitations to obtain tunable directional emission patterns and collective dissipative couplings when placing judiciously additional atoms nearby the atomic array. For doing that, we first characterize the optimal square array geometry to obtain  directional emission patterns. Then, we characterize the best atomic positions to couple efficiently to the subradiant metasurface excitations and provide several improvement strategies based on entangled atomic clusters or bilayers. Afterward, we also show how the directionality of the emission pattern can be controlled through the relative dipole orientation between the auxiliary atoms and the one of the array. Finally, we benchmark how these directional emission patterns translate into to collective, anisotropic dissipative couplings between the auxiliary atoms by studying the lifetime modification of atomic entangled states.
	\end{abstract}

	\maketitle
	
	
	The modification of atomic radiation by the presence of other atoms has been a very active of area in quantum optics since the seminal work by Dicke~\cite{dicke54a}. There, he showed that an atomic ensemble confined within a volume smaller than their optical wavelength ($\lambda_0$) emits photons with a collectively enhanced decay rate~\cite{gross82a} due to the photon-mediated interactions appearing between them~\cite{lehmberg70a,lehmberg70b}. With the advent of optical lattices~\cite{grimm00a,bloch08a}, the focus expanded to atomic arrays. Few works considered first the modification of the photonic energy dispersion for arrays with interatomic distances $d\sim \lambda_0$~\cite{Deutsch1995a,VanCoevorden1996a,Klugkist2006a,Antezza2009b,Antezza2009c,Antezza2013a,Bartolo2014b,Bartolo2014c}. However, the interest in the field exploded by studying the properties of (deeply) subwavelength arrays, that is, when $d<(\ll)\lambda_0$~\cite{porras08a,Scully2015,asenjogarcia17a,Asenjo-Garcia2019a,Zhang2019b,Bettles2016b,Wang2018,Bettles2016c,Bettles2020a,shahmoon17a,perczel17a,Perczel2017,bettles17a,Wild2018a,Glaetzle2017,Grankin2018a,Guimond2019b,Poddubny2020,Moreno-Cardoner2019a,Bekenstein2020,Alaee2020,Masson2020,Patti2021,Brechtelsbauer2020}. For such distances, interference leads to collective atomic responses very different from their individual one, like in metamaterials~\cite{Wang2016a}, and which can be harnessed to improve photon-storage fidelities~\cite{asenjogarcia17a} and quantum registers~\cite{Glaetzle2017}, to generate multiphoton states~\cite{Bekenstein2020}, or to obtain chiral~\cite{Grankin2018a} or magnetic~\cite{Alaee2020} light-matter interfaces. These prospects have placed such ``quantum metamaterials"~\cite{Bekenstein2020} at the spotlight, triggering several experiments~\cite{Glicenstein2020,Rui2020}.
	
	One of the most remarkable features of these systems is that they host subradiant excitations that propagate confined within them~\cite{asenjogarcia17a,Asenjo-Garcia2019a,Zhang2019b,Bettles2016b,Wang2018,Bettles2016c,Bettles2020a,shahmoon17a,perczel17a,Perczel2017,bettles17a,Wild2018a} with very long lifetimes. These subradiant excitations display nontrivial energy dispersions, like the photons propagating in photonic crystals~\cite{joannopoulos97a}, which can be tuned modifying the array configuration. This is why recent works~\cite{Masson2020,Patti2021,Brechtelsbauer2020} have pointed out these quantum metamaterials as a platform for exploring the physics of atoms coupled to photonic crystals~\cite{Chang2018}. Compared to nanophotonics, these systems (i) do not require complicated trapping schemes to place atoms nearby dielectrics~\cite{thompson13a,goban13a,Chang2019,Burgers2019,Luan2020a,Beguin2020}; (ii) energy dispersions of guided modes can be modified by optical means; and (iii) different from the guided modes in photonic crystals, these subradiant modes interact~\cite{Masson2020}, which can be used to induce gates~\cite{Rusconi2021}. So far, these initial works~\cite{Masson2020,Patti2021,Brechtelsbauer2020} have mostly considered the emergence of band-gap-mediated, coherent interactions, however, the possibilities are much richer~\cite{Bello2019,Leonforte2020b,DeBernardis2021,Garcia-Elcano2020,Garcia-Elcano2021,bravo12a,Gonzalez-Tudela2018,Perczel2020a,galve17a,Gonzalez-Tudela2017b,Gonzalez-Tudela2017a,Yu2019,Gonzalez-Tudela2018a}. 
	
	One of these exciting possibilities is the generation of anisotropic dissipative couplings between emitters when energetically tuned to van Hove singularities~\cite{galve17a,Gonzalez-Tudela2017b,Gonzalez-Tudela2017a,Yu2019,Gonzalez-Tudela2018a}. Despite its incoherent nature, such directional couplings lead to the formation of bound states in the continuum~\cite{hsu16a,Feiguin2020,Gonzalez-Tudela2017b,Gonzalez-Tudela2017a}, which can be instrumental to design quantum gates~\cite{Paulisch2016,kockum18a}. These initial works~\cite{galve17a,Gonzalez-Tudela2017b,Gonzalez-Tudela2017a} were based on simplified models, which neglected the coupling to free-space and polarization effects. Here, we propose a realistic quantum metasurface where these phenomena can be observed and controlled through the relative orientation and position of the impurity-array atomic dipoles. In addition, we show how the emission into the subradiant modes can be increased using entangled clusters or bilayer metasurfaces. We characterize these phenomena first with a single emitter, studying its emission, and then with many, characterizing the lifetime modification of entangled states, i.e., the signature of super- and subradiance~\cite{dicke54a}.
	
	\begin{figure}[!htbp]
		\centering
		\includegraphics[width=\linewidth]{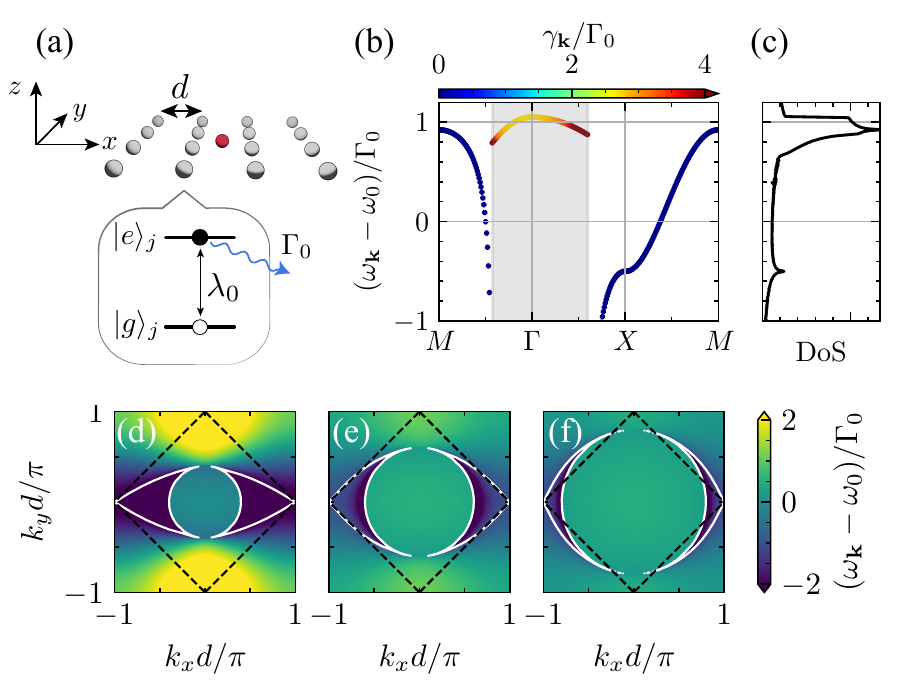}
		\caption{(a) Impurity atoms (red) are placed near a square atomic array (gray) with lattice constant $d$. Each array and impurity atom has a single optical transition of wavelength $\lambda_{0/a}$ and a free-space decay rate of $\Gamma_{0/a}$. (b) Band structure for an array with in-plane polarization $\hat{\boldsymbol{\wp}}_0 = \hat{\mathbf{e}}_y$ and $d/\lambda_0 = 0.3$. Color scale represents the collective decay rate, and the gray shadow region denotes the light cone. (c) Density of state in arbitrary units for the band structure shown in (b). (d)-(f) Color maps of the array band structure with (d) $d/\lambda_0=0.2$, (e) $d/\lambda_0=0.3$ and (f) $d/\lambda_0=0.4$. Equipotentials for the energy $\omega_{\mathbf{k} = X}$ for the metasurface (solid white) compared to the nearest-neighbor model~\cite{galve17a,Gonzalez-Tudela2017b,Gonzalez-Tudela2017a} (dashed black).
		}
		\label{fig:1-scheme}
	\end{figure}
	
	The setup we consider in this Letter is depicted in Fig.~\ref{fig:1-scheme}(a): auxiliary atoms are placed near a square atomic array with interatomic distance $d$. For simplicity, we consider atomic systems with a single optical transition ($e-g$) of frequency $\omega_{a/0}=k_{a/0}c=2\pi c/\lambda_{a/0}$ and polarization $\boldsymbol{\wp}_{a/0}$ for the impurity/array atoms, respectively. Here, we focus on the situation in which the array dipoles are oriented in-plane (see Supplemental Material \cite{SupMatDirectional} for the out-of-plane case), e.g., fixing $\boldsymbol{\wp}_0=\hat{\bf{e}}_y$. We leave $\boldsymbol{\wp}_a$ as a free parameter, which, as we see below, will allow us to tune the emergent behavior. 
	
	Let us first characterize the properties of the metasurface. Its dynamics can be described within an stochastic wave function approach through the following non-Hermitian Hamiltonian~\cite{Masson2020}:
	\begin{equation}
		\frac{H_{m}}{\hbar} = \sum_{j=1}^N\left(\omega_0-i\frac{\Gamma_0}{2}\right) \sigma^j_{ee}+\sum_{i\neq j=1}^N\left(J_{ij}-i\frac{\Gamma_{ij}}{2}\right)\sigma_{eg}^i\sigma_{ge}^j\,,
		\label{eq:H_eff}
	\end{equation}
	where the atom $j$-th is located at $\rr_j$, $\Gamma_0=|\boldsymbol{\wp}_0|^2\omega_0^3/(3\pi\hbar c^3)$ is the individual free-space decay rate, and $\sigma_{\alpha\beta}^j=\ket{\alpha}_j\bra{\beta}$ the dipole operators. The coherent ($J_{ij}$) and incoherent ($\Gamma_{ij}$) emitter interactions are given by the vacuum's Green's Function~\cite{asenjogarcia17a} $\mathbf{G}_0(\rr_i-\rr_j)$,
	\begin{equation}
		J_{ij}-i\frac{\Gamma_{ij}}{2}=-\frac{3\pi \Gamma_0}{\omega_0}\hat{\boldsymbol{\wp}}^*_{i} \cdot\mathbf{G}_0(\rr_i-\rr_j)\cdot\hat{\boldsymbol{\wp}}_{j}
		\label{eq:Green_tensora}
	\end{equation}
	where $\hat{\boldsymbol{\wp}}_i=\boldsymbol{\wp}_{i}/|\boldsymbol{\wp}_i|\equiv \hat{\boldsymbol{\wp}}_0$. 
	%
	In the single-excitation subspace and infinite size limit, the eigenstates of the Hamiltonian $H_{m}$ are Bloch functions $S^\dagger_\kk=\frac{1}{\sqrt{N}}\sum_j \sigma^j_{eg} e^{i\kk\cdot\rr_j}$, where $\kk=(k_x,k_y)\in [-\pi/d,\pi/d]^{\otimes 2}$, and their (complex) eigenenergies read
	\begin{align}
		\omega_\kk-i\frac{\gamma_\kk}{2}=\omega_0-\frac{3\pi \Gamma_0}{k_0} \hat{\boldsymbol{\wp}}^*_0 \cdot\tilde{\bf{G}}_0(\kk)\cdot\hat{\boldsymbol{\wp}}_0\,,
	\end{align}
	where $\tilde{\bf{G}}_0(\kk)=\sum_j e^{-i\kk\cdot\rr_j} \bf{G}_0(\rr_j)$ is the discrete Fourier transform of the free-space tensor. In Fig.~\ref{fig:1-scheme}(b) we plot the energy dispersion $\omega_\kk$ and their associated imaginary part $\gamma_\kk$ (in color scale) along a path of the Brillouin zone and for an array with $d/\lambda_0=0.3$~\cite{SupMatDirectional}. As expected for such distances, interference effects lead to the sub(super) radiant character [$\gamma_\kk< (>) \Gamma_0$] of the eigenstates outside (within) the light cone. Its energy dispersion $\omega_\kk$ features a saddle point at the X point, which leads to a van Hove singularity in the density of states at its energy, see Fig.~\ref{fig:1-scheme}(c). This singularity also appears in the nearest-neighbor model, and it is where the anisotropic emission and collective interactions for resonant emitters were predicted~\cite{galve17a,Gonzalez-Tudela2017b,Gonzalez-Tudela2017a}. In that case, however, the saddle point is accompanied by straight isofrequencies, i.e., $k_x \pm (\mp) k_y=\pm \pi/d$, important for the emission directionality. The long-range nature of the photon-mediated interactions in free space~\cite{asenjogarcia17a} modify this behavior. This is shown explicitly in Figs.~\ref{fig:1-scheme}(d)-\ref{fig:1-scheme}(f), where we plot $\omega_\kk$ and the corresponding isofrequency line at the X point for several $d/\lambda_0$. By doing a systematic analysis~\cite{SupMatDirectional}, we find that $d/\lambda_0\approx 0.3$ leads to an optimal performance for in-plane polarized modes because it maximizes the isofrequency straightness, the density of states at that energy, and, as we see next, its tunability. Such subwavelength regime can be obtained by using a different optical transition for trapping the atoms than for mediating the interactions, as already done in Ref.~\cite{Rui2020}. Thus, alkaline-Earth atoms look particularly suitable since they feature optical transitions from the near ultraviolet to the infrared range~\cite{ludlow15a,covey19a}.

	\begin{figure}[tb]
		\centering
		\includegraphics[width=\linewidth]{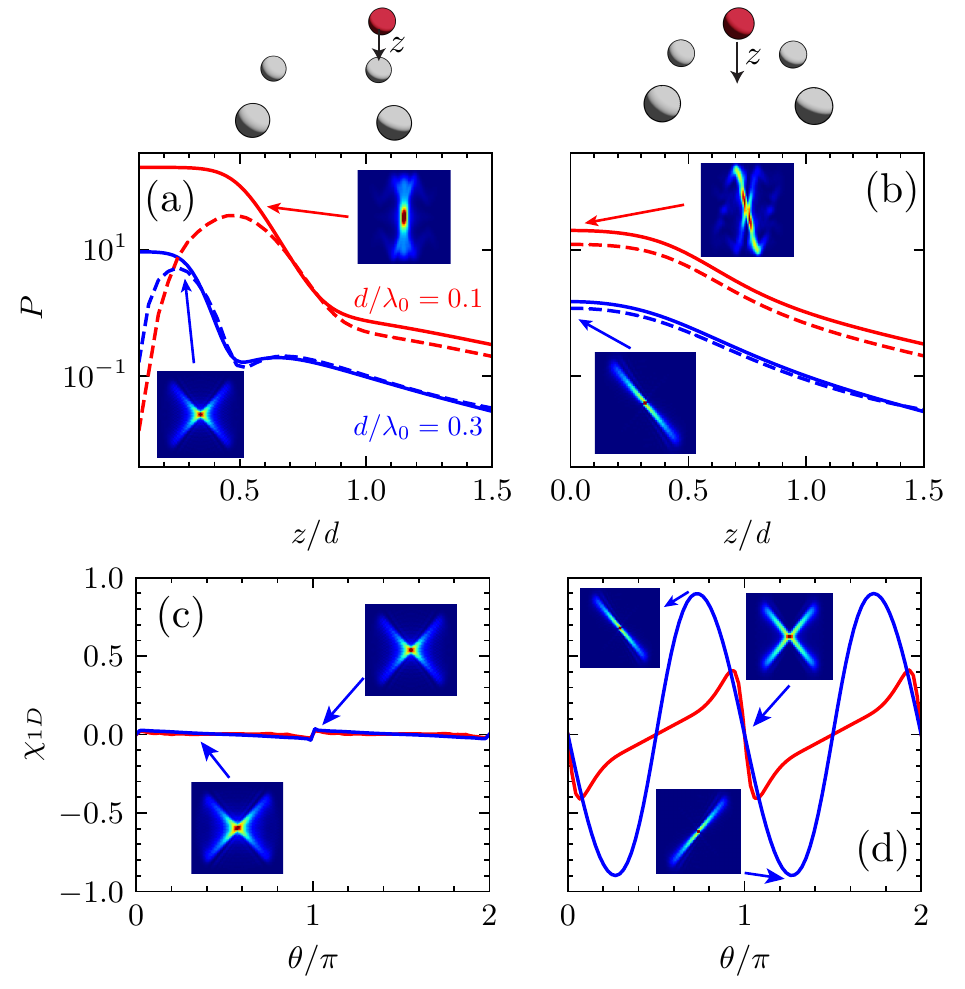}
		\caption{(a),(b) Semianalytical (solid lines) and numerical (dashed lines) Purcell factors as function of the vertical distance of the emitter when it is located (a) on top of an atom, (b) in the center of a plaquette. The atomic array has in-plane polarization $\hat{\boldsymbol{\wp}}_0 = \hat{\mathbf{e}}_y$, while the impurity atom has $\hat{\boldsymbol{\wp}}_a = (\hat{\mathbf{e}}_x + \hat{\mathbf{e}}_y) / \sqrt{2}$ and a individual decay rate $\Gamma_a = 0.002\Gamma_0$. Numerical calculations have been obtained with a $50\times 50$ dipole array. (c), (d) Directionality parameter $\chi_{\mathrm{1D}}$ as a function of the relative orientation of the impurity-array dipoles $\theta$, with the emitter (c) on top of a dipole at $z=0.5 d$ ($z=0.3d$) for $\lambda_0/d=0.1$ ($\lambda_0/d=0.3$) and (d) in the center of a plaquette with $z=0$. Insets show the emission patterns in real space at the points denoted by the arrows.  Horizontal and vertical axes represent the $x$ and $y$ directions, respectively.}
		\label{fig:2-purcelldirectional}
	\end{figure}

	Let us now consider the effect of placing an impurity atom near the array at position $\rr_a$. The dynamics of the combined system is described by the Hamiltonian $H=H_a+H_{am}+H_m$, where~\footnote{Note, here we use the assumption commonly used in the literature that the free-space Green's function does not vary significantly for the energy ranges around $\omega_a,\omega_0$.}
	
	\begin{subequations}
		\begin{align}
			\frac{H_a}{\hbar} &= \left(\omega_a-i\frac{\Gamma_a}{2}\right)\sigma_{ee}^a,\\
			\frac{H_{am}}{\hbar} &= \sum_{i=1}^{N}\left(J_{ai} - i\frac{\sqrt{\Gamma_0\Gamma_a}}{2}\right)\sigma_{eg}^a\sigma_{ge}^i + (i\leftrightarrow a).
		\end{align}
	\end{subequations}

	The impurity atom can either be the same atom or a different isotope, and if one can isolate $\Lambda$ scheme in its level structure, its frequency and linewidth can be controlled with Raman-assisted transitions (see~\cite{porras08a,Brechtelsbauer2020,SupMatDirectional}). To characterize how well the impurity atoms couple to the guided modes of the metasurface, we calculate the metasurface Purcell factor ($P$), that is, the ratio between the decay rate into subradiant modes ($\Gamma_{m}$) compared to free-space modes ($\Gamma'$). We do it using two complementary approaches. First, the semianalytical approach developed in Ref.~\cite{asenjogarcia17a},
	\begin{equation}
		P_a = \frac{\frac{9d^2}{2k_a^2}\Im\left(\iint _{\abs{\mathbf{k}}> k_0}d^2\mathbf{k}\frac{\hat{\boldsymbol{\wp}}^*_a \boldsymbol{\alpha}_\mathbf{k}(\mathbf{r}_a)\otimes \boldsymbol{\beta}_\mathbf{k}(\mathbf{r}_a)\hat{\boldsymbol{\wp}}_a}{(\omega_a-\omega_\mathbf{k})/\Gamma_0}\right)}{1 + \frac{9d^2}{2k_0^2}\Im\left(\iint _{\abs{\mathbf{k}}\leq k_0}d^2\mathbf{k}\frac{\hat{\boldsymbol{\wp}}^*_a \boldsymbol{\alpha}_\mathbf{k}(\mathbf{r}_a)\otimes \boldsymbol{\beta}_\mathbf{k}(\mathbf{r}_a)\hat{\boldsymbol{\wp}}_a}{(\omega_a-\omega_\mathbf{k})/\Gamma_0}\right)},
		\label{eq:ratio}
	\end{equation}
	where $\boldsymbol{\alpha}_\mathbf{k}(\mathbf{r}_a)$ and $\boldsymbol{\beta}_\mathbf{k}(\mathbf{r}_a)$ are the field eigenmodes evaluated at the impurity atom position $\rr_a$~\cite{SupMatDirectional},
	\begin{subequations}
		\begin{align}
			\boldsymbol{\alpha}_\mathbf{k}(\mathbf{r}) &= \sum_{j=1}^N\mathbf{G}_0(\mathbf{r}, \mathbf{r}_j, \omega_0)\cdot \hat{
				\boldsymbol{\wp}}_0e^{i\mathbf{r}_j\cdot\mathbf{k}},\\
			\boldsymbol{\beta}_\mathbf{k}(\mathbf{r}) &= \sum_{j=1}^N\hat{
				\boldsymbol{\wp}}_0^*\mathbf{G}_0(\mathbf{r}_j, \mathbf{r}, \omega_0)\cdot e^{-i\mathbf{r}_j\cdot\mathbf{k}}.
		\end{align}
	\end{subequations}

	This expression is obtained under the Born-Markov approximation, which neglects retardation within the array and assumes $\Gamma_{m}$ is much smaller than the bandwidth of $\omega_\kk$. To avoid relying on this assumption and to get a better picture of real experiments, we alternatively calculate $P$ by solving exactly the dynamics assuming an initially excited impurity atom, i.e., $\sigma^a_{eg}\ket{\mathrm{vac}}$. Since $H$ is excitation preserving, the system wave function at any time reads
	\begin{align}
		\ket{\Psi(t)}=\left(C_a(t)\sigma^a_{eg}+\sum_{j=1}^N C_{\rr_j}(t)\sigma_{eg}^j\right)\ket{\mathrm{vac}}\,.
	\end{align}
	
	From this wave function one can obtain a numerical estimation of the Purcell factor~\cite{SupMatDirectional}, which we label as $P_n$ and which takes into account non-Markovian effects. Additionally, plotting $|C_{\rr_j}(t)|$, one obtains the spatial emission pattern, which will be directly related to how the impurity atoms interact among them. 
	
	In Figs.~\ref{fig:2-purcelldirectional}(a) and \ref{fig:2-purcelldirectional}(b), we plot the dependence of the Purcell factor on the vertical distance $z$ for emitters placed above an atom or at the center of the unit cell, respectively, as well as its emission pattern (insets). We compare both the semianalytical $P_a$ (solid lines) and the numerical approach $P_n$ (dashed lines) for two different distances, i.e., $d/\lambda_0=0.1$ and $0.3$ in red and blue, respectively. Closer interatomic distances of the impurity atom to the metasurface lead to larger Purcell factors, although at the expense of losing the cross-directional emission, as expected from Figs.~\ref{fig:1-scheme}(d)-\ref{fig:1-scheme}(f). In addition, it is also at these small $z$ regions where we see the larger deviations between the semianalytical and numerical Purcell factors. These differences can be attributed to strong deviations from the Markovian behavior~\cite{SupMatDirectional}, where $P_a$ is not expected to work, and can be attenuated by reducing $\Gamma_a$, e.g., with a Raman transition. Apart from this deviation, another important difference of placing the impurity atom exactly above a metasurface atom [Fig.~\ref{fig:2-purcelldirectional}(a)] or at the center of the unit cell [Fig.~\ref{fig:2-purcelldirectional}(b)] is the tunability of the cross-directional emission shown in the inset of both panels. In particular, we can show that changing the relative orientation between   the lattice and the impurity atom $\theta=\arccos\left(\hat{\boldsymbol{\wp}}_0\cdot\hat{\boldsymbol{\wp}}_a\right)$ cancels the emission along one of the directions in the former case, but not in the latter. To characterize qualitatively this tunability, we define a directional parameter~\cite{SupMatDirectional},
	
	\begin{equation}
		\chi_{\mathrm{1D}} = \sum_{\abs{\mathbf{r}_j}\sim R}\tilde{\mathcal{P}}_j\cos[2(\theta_j - \theta_{\max})],
	\end{equation}
	where $\tilde{\mathcal{P}_j}$ is the cumulative population of the dipoles located near a circle of radius $R$ centered on the emitter, and $\theta_j$ is its angle with respect to the $x$ axis. The maximum cumulative population is located at the angle $\theta_{\max}$. Cumulative population is renormalized so $\sum_{\abs{\mathbf{r}_j}\sim R} \tilde{\mathcal{P}}_j=1$. With this definition, $\chi_{\mathrm{1D}}=1$ when the emission is purely one-dimensional, whereas $\chi_{\mathrm{1D}}\approx 0$ when it becomes isotropic or emits in two orthogonal directions. In Figs.~\ref{fig:2-purcelldirectional}(c) and \ref{fig:2-purcelldirectional}(d) we plot $\chi_{\mathrm{1D}}(\theta)$ for the impurity positions of panels Figs.~\ref{fig:2-purcelldirectional}(a) and \ref{fig:2-purcelldirectional}(b), respectively, showing that $\chi_{\mathrm{1D}}(\theta)\approx 1$ for certain $\theta$ with impurities at the center of the unit cell, whereas $\chi_{\mathrm{1D}}(\theta)\approx 0$ at all $\theta$ for the other case.

	\begin{figure}[!htbp]
		\centering
		\includegraphics[width=0.9\linewidth]{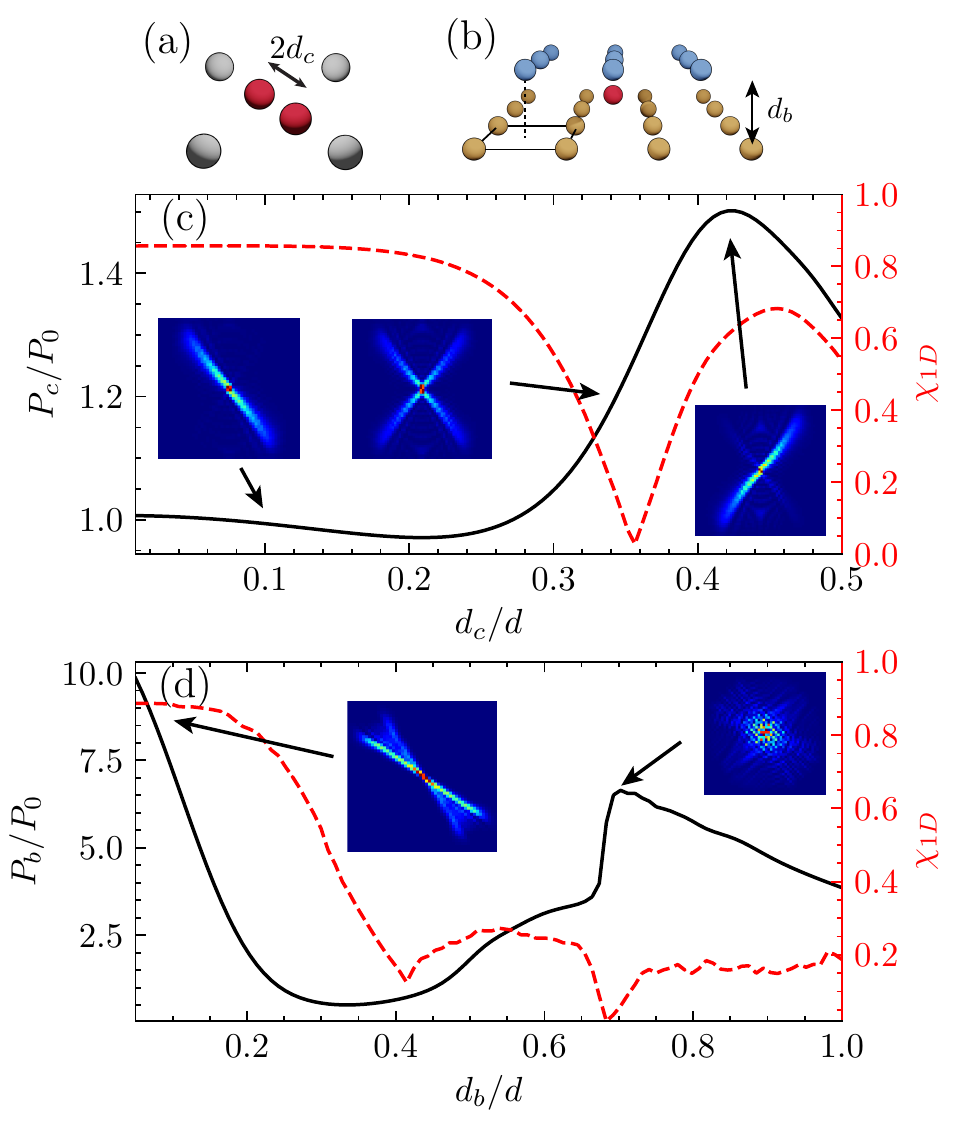}
		\caption{(a),(b) Entangled cluster and bilayer schemes. (c),(d) Purcell factor $P$ (solid black lines, left axis) and directionality $\chi_{\mathrm{1D}}$ (dashed red lines, right axis) for (c)  a single cluster placed near the simple metasurface and (d) a single emitter placed within a bilayer metasurface. (c) The impurities within the cluster are placed at  $\mathbf{r}_{1,2} = (\pm d_c / \sqrt{2}, \mp d_c/\sqrt{2}, 0)$, with a polarization of $\hat{\boldsymbol{\wp}}_a = (\hat{\mathbf{e}}_x + \hat{\mathbf{e}}_y) / \sqrt{2}$ and initialized in a symmetric state. The array has a polarization $\hat{\boldsymbol{\wp}}_0=\hat{\mathbf{e}}_y$. (d) The two layers are shifted by $(0.5 d, 0.5d)$, and the emitter is located in the middle of both layers in $\mathbf{r}_a = (0, 0.5d, d_b/2)$. The dipoles of the layers have a polarization $\hat{\boldsymbol{\wp}}_0  = (\hat{\mathbf{e}}_x + \hat{\mathbf{e}}_y) / \sqrt{2}$, while for the emitters $\hat{\boldsymbol{\wp}}_a =\hat{\mathbf{e}}_y$. In both cases, $d/\lambda_0 = 0.3$ and emitters are resonant with the $X$ modes. Insets show the emission patterns in real space ($x$ and $y$ directions) at the configuration denoted by the arrows. Horizontal and vertical axes represent the $x$ and $y$ directions, respectively.}
		\label{fig:3-Strategies}
	\end{figure}
	
	In Fig.~\ref{fig:2-purcelldirectional}, we see how, for the most tunable situation ($d/\lambda_0=0.3$), the Purcell factor is limited to $P\approx 1$. Now, we explore two possible strategies to boost $P$ while still preserving the possibility of tuning the directional emission patterns. The first strategy consists of placing pairs of emitters separated a distance $d_p$ and prepared in a given entangled state, e.g., $\ket{\Psi_c}=(\ket{e_{a_1}g_{a_2}}\pm\ket{g_{a_1}e_{a_2}})/\sqrt{2}$. This can be done, e.g., using dynamical optical tweezers and local spin exchange~\cite{Kaufman2015} or Rydberg interactions~\cite{Wilk2010}. The intuition is that the interference between the atomic emission within the cluster can lead to a cancellation of free-space emission and ultimately boost $P$. This occurs for the antisymmetric case, obtaining $P\sim 10$, but at the price of losing the tunability of the directional emission (see Supplemental Material~\cite{SupMatDirectional}). Since for this Letter, we are interested in keeping the tunability, in Fig.~\ref{fig:3-Strategies}(c) we compare the numerically obtained $P_c$ for the symmetric cluster with the individual one $P_0$ for a dipole in the center of a plaquette at $z=0$ as a function of the cluster distance $d_c$. Like this, one can get a moderate improvement $P_c/P_0\approx 1.5$ for $d_c\approx 0.45d$, but keeping the directional emission. Besides, we observe an effective rotation of the emission as $d_c$ increases, which shows that $d_c$ can be used as a tuning knob by itself. The other strategy consists of changing the metasurface structure by a bilayer one with separation $d_b$, see scheme in Fig.~\ref{fig:3-Strategies}(b). This structure also features straight isofrequencies~\cite{SupMatDirectional}, where one can energetically tune the impurity atom to obtain directional emission patterns. In Fig.~\ref{fig:3-Strategies}(d), we numerically obtain $P_b$ for the bilayer system and show how it can reach $P_b/P_0\approx 7$ for either small $d_b$ distances, keeping the 1D tunability, or for $d_b\approx 0.7$ but losing the 1D character of the emission.
	
	\begin{figure}
		\centering
		\includegraphics[width=0.85\linewidth]{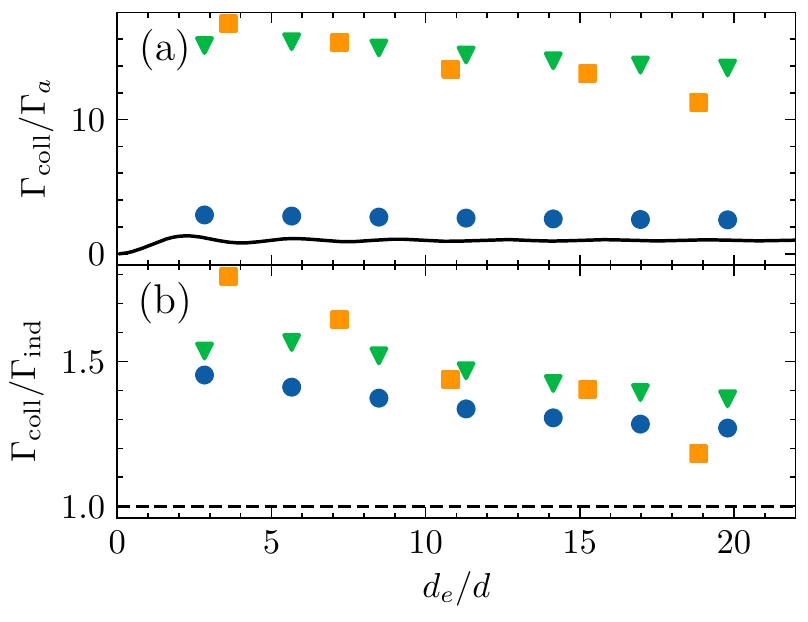}
		\caption{(a),(b) Collective decay rates with the different strategies (a) normalized to the individual emitter decay rate in free space and (b) normalized to the individual decay rate at each configuration, for a monolayer with two emitters (blue circles), a monolayer with two clusters (green triangles), and a bilayer with two emitters (orange squares). The black solid line represents the case with no metasurface. Cluster impurities distance $d_c = 0.45d$, bilayer distance $d_b=0.1d$. Other parameters are the same as those in Fig.~\ref{fig:3-Strategies}.}
		\label{fig:4-collective}
	\end{figure}

	After having characterized the single impurity coupling, let us consider how the directional emission patterns translate into collective dissipative interactions when more impurity atoms couple to the metasurface. One typical signature of these dissipative couplings is the lifetime renormalization of entangled atomic states~\cite{dicke54a}. This means that if an individual system decays with a rate $\Gamma_\mathrm{ind}$, an entangled pair features a collective enhancement (decrease) of such decay rate $\Gamma_\mathrm{coll}> (<)\Gamma_\mathrm{ind}$ depending on whether it is a super(sub) radiant configuration. In Fig.~\ref{fig:4-collective}, we extract this collective enhancement $\Gamma_\mathrm{coll}$ as a function of the distance between impurities $d_e/d$ through a numerical fitting of the dynamics for the three different configurations explored in Figs.~\ref{fig:2-purcelldirectional}-\ref{fig:3-Strategies}: namely, (i) for a pair of impurity atoms near the single-layer array (blue dots); (ii) for a pair of entangled atomic clusters near the single-layer array (green dots); and (iii) for a pair of impurity atoms within a shifted bilayer array (orange dots). In Fig.~\ref{fig:4-collective}(a), we plot $\Gamma_\mathrm{coll}$ normalized to the impurity atom free-space individual decay rate $\Gamma_a$. There we observe how, indeed, all strategies provide an improvement of collective effects compared to the case with no metasurface. To better compare the intrinsic collective dynamics induced in the different systems, in Fig.~\ref{fig:4-collective}(b) we plot the same configurations, but normalizing $\Gamma_\mathrm{coll}$ to the individual decay rate $\Gamma_\mathrm{ind}$ in each configuration. There we observe how the bilayer enhances better collective effects at small distances due to its imperfect directionality, see Fig.~\ref{fig:3-Strategies}(d), whereas the entangled cluster performs better at larger distances.
	
	Summing up, we show how to obtain strongly directional emission patterns by placing atoms near quantum metasurfaces. We also study several strategies to achieve more efficient couplings between the impurity atoms and the directional subradiant modes of the metasurfaces based on entangled clusters or bilayer systems. Finally, we also show how these directional emission patterns translate into collective dissipative couplings when more impurity atoms couple to the metasurface. This shows the potential of quantum metasurfaces to induce nontrivial collective dissipative effects resulting from the interplay of interference and unconventional band structures. Our results can be of interest as well for subwavelength exciton arrays in 2D materials~\cite{Reserbat-Plantey2021}. An interesting outlook is to extend this analysis to the case topological band-structure models~\cite{perczel17a,Gonzalez-Tudela2018,Perczel2020a}.
	
	\begin{acknowledgments}
		A.G.-T. acknowledges financial support from the Proyecto Sinérgico CAM 2020 Y2020/TCS-6545 (NanoQuCo-CM), the CSIC Research Platform on Quantum Technologies PTI-001 and from Spanish project PGC2018-094792-B-100(MCIU/AEI/FEDER, EU). D.F.F acknowledges support from the CSIC JAE Intro Research program JAEINT-20-01823/EX-0330.
	\end{acknowledgments}\;
	
	\textit{Note added.}-Recently, two works exploring similar ideas were published~\cite{Castells-Graells2021a,Zundel2021}. The code to reproduce the figures of this Letter can be found in \cite{CodeZenodo}.
		
	
	\newpage
	\begin{widetext}
		
		\begin{center}
			\textbf{\large Supplemental Material: Tunable Directional Emission and Collective Dissipation with Quantum Metasurfaces  \\}
		\end{center}
		\setcounter{equation}{0}
		\setcounter{figure}{0}
		\makeatletter
		
		\renewcommand{\thefigure}{SM\arabic{figure}}
		\renewcommand{\thesection}{SM\arabic{section}}  
		\renewcommand{\theequation}{SM\arabic{equation}}
		
		This Supplementary Material is divided as follows: in~\ref{secSM:bandsinglelayer} we explain in detail how to calculate the emergent band-structure of the single-layer metasurface, including the results for the in-plane and out-of-plane polarized modes; in Section~\ref{secSM:impuritycoupling} we explain how to characterize impurity atom-metasurface coupling following both the semi-analytical approach of Ref.~\cite{asenjogarcia17a}, and the numerical strategy that we design for this manuscript; in Section~\ref{secSM:tunabledirectional} we provide more details on how to tune the directional emission; in Section~\ref{secSM:improvement} we explain the improvement strategies developed in the main part of the manuscript based on entangled clusters and bilayer structures; in Section~\ref{secSM:collective} we explain the different configurations used to explore the collective superradiant dynamics of Fig. 4 of the main text; in Section~\ref{secSM:triangular}, we show that that similar phenomena can be found in other geometries by studying the emission pattern in a triangular geometry metasurface; finally, in Section~\ref{secSM:experimental} we review the different experimental ingredients required to implement these ideas.
		
		\section{Band-structure of the single layer configuration~\label{secSM:bandsinglelayer}}
		
		In this section, we are interested in characterizing completely the eigen-modes and eigen-energies of the sub-wavelength atomic array of Fig. 1(a) of the main manuscript. This quantum metasurface has a square geometry with primitive vectors $\aa_{1/2}=d\hat{\bf{e}}_{x/y}$, being $d$ the inter-atomic distance. The array has a total $N=N_l^2$ number of atoms, where $N_l$ denotes the linear size of the system.  Thus, the position of the $j$-th atom, $\rr_j$ can be described by two integer indices $n_i=0,1,\dots,N_l$ such that $\rr_j=n_1 \aa_{1}+n_2\aa_2$. For simplicity, we assume a two-level system description of the atoms. This means that the atoms only have a single optical transition between an excited and ground state denoted by $e$ and $g$, respectively, with frequency $\omega_0=c k_0=c 2\pi/\lambda_0$ and dipole polarization $\boldsymbol{\wp}_{j}$. In the main text, we only characterize the physics of the in-plane modes assuming $\hat{\boldsymbol{\wp}}_{j}\equiv\hat{\boldsymbol{\wp}}_{0}=\hat{\bf{e}}_{y}$. For completeness, here we will also calculate the emergent band-structure of the out-of-plane modes, $\boldsymbol{\wp}_{0}=\hat{\bf{e}}_{z}$, as well as the case of a circularly-polarized transition $\hat{\boldsymbol{\wp}}_{0}=(\hat{\bf{e}}_{x} + i\hat{\bf{e}}_{y})/\sqrt{2}$.
		
		Irrespective of the polarization chosen, the dynamics of the subwavelength atomic array is described within a master equation formalism through the following Born-Markov master equation~\cite{lehmberg70a,lehmberg70b}:
		
		\begin{align}
			\hbar\frac{d \rho}{dt}=i(\rho H^\dagger_{m}-H_m\rho)+\sum_{i\neq j}\Gamma_{ij}\sigma_{ge}^i\rho \sigma_{eg}^j\,,\label{eqSM:mequation}
		\end{align}
		
		where $H_m$ is an effective non-Hermitian Hamiltonian given by
		
		\begin{equation}
			H_{m}/\hbar = \sum_{j=1}^N\left(\omega_0-i\frac{\Gamma_0}{2}\right) \sigma^j_{ee}+\sum_{i\neq j=1}^N\left(J_{ij}-i\frac{\Gamma_{ij}}{2}\right)\sigma_{eg}^i\sigma_{eg}^j\,,
			\label{eqSM:H_eff}
		\end{equation}
		
		where $j$ is a index running over the number of atoms of the metasurface, $\Gamma_0=|\boldsymbol{\wp}_0|^2\omega_0^3/(3\pi\hbar c^3)$ is the individual free-space decay rate, and $\sigma_{\alpha\beta}^j=\ket{\alpha}_j\bra{\beta}$ the spin-dipole operator of the $j$-th atom. The coherent ($J_{ij}$) and incoherent ($\Gamma_{ij}$) photon-mediated interactions among emitters are given by the vacuum's Green Function~\cite{lehmberg70a,lehmberg70b} $\mathbf{G}_0(\rr_i-\rr_j)$:
		
		\begin{equation}
			J_{ij}-i\frac{\Gamma_{ij}}{2}=-\frac{3\pi \Gamma_0}{\omega_0}\hat{\boldsymbol{\wp}}^*_{i} \cdot\mathbf{G}_0(\rr_i-\rr_j)\cdot\hat{\boldsymbol{\wp}}_{j}
			\label{eqSM:Green_tensora}
		\end{equation}
		
		where:
		\begin{equation}
			\mathbf{G}_0(\mathbf{r})= \frac{e^{ik_0r}}{4\pi k_0^2r^3}\left[(k_0^2r^2+ik_0r-1)\mathds{1}_{3\times3}
			+(-k_0^2r^2-3ik_0r+3)\frac{\mathbf{r}\otimes\mathbf{r}}{r^2}\right],
			\label{eqSM:Green_tensor1}
		\end{equation}
		with $r = |\mathbf{r}|$. Note these expressions of Eqs.~\eqref{eqSM:H_eff}-\eqref{eqSM:Green_tensor1} have already implicit the Born-Markov approximations~\cite{lehmberg70a,lehmberg70b}, which assumes that the frequency dependence of $\mathbf{G}(\mathbf{r})$ enters directly through the atomic frequency $\omega_0$.
		
		In the thermodynamic limit ($N\rightarrow \infty$) the momentum $\kk$ is a good quantum number. This means that the Hamiltonian $H_m$, in the single-excitation subspace, can be diagonalized using the Bloch functions $S^\dagger_\kk=\frac{1}{\sqrt{N}}\sum_j \sigma^j_{eg} e^{i\kk\cdot\rr_j}$ as follows:
		
		\begin{align}
			\frac{H_m}{\hbar}=\sum_{\kk}\left(\omega_\kk-i\frac{\gamma_\kk}{2}\right) S^\dagger_\kk S_\kk\,,
		\end{align}
		
		where $\kk=(k_x,k_y)$ takes values within the first Brillouin zone of the structure, that is $k_{x,y}\in [-\pi/d,\pi/d]$. Each $\kk$ mode has then associated both a real ($\omega_\kk$) and imaginary ($\gamma_\kk$) contribution to their energy given by
		
		\begin{align}
			\omega_\kk-i\frac{\gamma_\kk}{2}=\omega_0-\frac{3\pi \Gamma_0}{k_0} \hat{\boldsymbol{\wp}}^*_{0} \cdot\tilde{\bf{G}}_0(\kk)\cdot\hat{\boldsymbol{\wp}}_{0}\,,\label{eqSM:omkgammak}
		\end{align}
		
		where $\tilde{\bf{G}}_0(\kk)=\sum_{j} e^{-i\kk\cdot\rr_j} \bf{G}_0(\rr_j)$ is the discrete Fourier transform of the free-space tensor. Note that due to the long-range nature of the photon-mediated interactions of $\bf{G}_0(\rr_j)$ (see Eq.\eqref{eqSM:Green_tensor1}), the sum in real space is slowly convergent. A method of avoiding that was proposed in Ref.~\cite{Perczel2017} which consists in performing the summations in momentum space. This can be done using Poisson's identity to write the sum in $\tilde{G}_0(\kk)$ as a summation over reciprocal lattice vectors:
		\begin{equation}
			\sum_{\mathbf{r}_n\neq0} e^{\mathbf{k}\cdot \mathbf{r}_n}\mathbf{G}(\mathbf{r}_n) = \frac{1}{\mathcal{A}}\sum_{\boldsymbol{\mathcal{G}}}\mathbf{g}(\boldsymbol{\mathcal{G}}-\mathbf{k})- \bf{G}(0),
			\label{eqSM:Poisson}
		\end{equation}
		
		where $\mathcal{A}$ is the area of the unit cell, and $\mathbf{g}$ is the Weyl decomposition of the Green tensor, which is given by
		
		\begin{equation}
			g_{\alpha\beta}(\mathbf{p}) = \int\frac{dp_z}{2\pi}\frac{1}{k_0^2}\frac{k_0^2\delta_{\alpha\beta}-p_\alpha p_\beta}{k_0^2-p^2}.\label{eqSM:weylg2}
		\end{equation}
		
		Note that in Eq.~\eqref{eqSM:Poisson} we separated the $\mathbf{r}_n=0$ component because it provides a overall energy shift and it is mathematically divergent. Physically, this divergence is smeared out due to two effects: first, the system is finite and thus, one does not actually deal with infinite summation terms; second, the atoms are not point-dipole emitters, but rather have a well-defined finite size given by its ground-state atomic wavefunction size, $a_{\text{ho}}$. It can be shown that both terms by themselves renormalize the divergences and result in physical band-structures. 
		
		For example, in Ref.~\cite{Perczel2017}, they obtained expressions for the renormalized Green tensor taking into account finite atomic size $a_{\text{ho}}$. This results into a convergent Green tensor at the origin:
		
		\begin{equation}
			\mathbf{\bf{G}}^0(0) = \frac{k_0}{6\pi}\left[\frac{\operatorname{erfi}(k_0a_\text{ho}/\sqrt{2})-i}{\exp((ka_\text{ho})^2/2)}+\frac{1/2-(k_0a_\text{ho})^2}{(\pi/2)^{1/2}(k_0a_\text{ho})^3}\right]\mathds{1}_{3\times 3},
		\end{equation}
		
		where $\operatorname{erfi}(z)$ is the imaginary error function. On top of that, they also obtained the renormalization of the rest of the components of the Green tensor:
		\begin{subequations}
			\begin{align}
				g^*_{xx}(\mathbf{p}_{x-y})&=(k_0^2-p_x^2)\mathcal{I}_0\,,\, g^*_{yy}(\mathbf{p}_{x-y})=(k_0^2-p_y^2)\mathcal{I}_0\,,\,
				g^*_{zz}(\mathbf{p_{x-y}})=(k_0^2\mathcal{I}_0-\mathcal{I}_2),\\
				g^*_{xy}(\mathbf{p}_{x-y})&=g^*_{yx}(\mathbf{p}_{x-y})=-p_xp_y\mathcal{I}_0,\\
				g^*_{xz}(\mathbf{p}_{x-y})&=g^*_{zx}(\mathbf{p}_{x-y})=	g^*_{yz}(\mathbf{p}_{x-y})=g^*_{zy}(\mathbf{p}_{x-y})=0\,,
			\end{align}
		\end{subequations}
		
		where we have defined the following expressions
		\begin{equation}
			\mathcal{I}_0(\mathbf{p}_{x-y})=\mathcal{C}\frac{\pi e^{-a_\text{ho}^2\Lambda^2/2}}{\Lambda}\left[-i+\operatorname{erfi}\left(\frac{a_\text{ho}\Lambda}{\sqrt{2}}\right)\right],
		\end{equation}
		\begin{equation}
			\mathcal{I}_2(\mathbf{p}_{x-y})=\Lambda^2\mathcal{I}_0-\mathcal{C}\frac{\sqrt{2\pi}}{a_\text{ho}},
		\end{equation}
		\begin{equation}
			\mathcal{C}(\mathbf{p}_{x-y})=\frac{1}{2\pi k_0^2}e^{-a_\text{ho}^2p^2/2},
		\end{equation}
		\begin{equation}
			\Lambda(\mathbf{p}_{x-y})=\sqrt{k_0^2-p^2}\geq 0.
		\end{equation}
		
		An alternative approach consists in directly diagonalizing the Hamiltonian $H_m$ of Eq.~\eqref{eqSM:H_eff} for a finite system. In that case, $\kk$ is not strictly a good quantum number, however, the eigenstates that are formally written as:
		\begin{equation}
			\ket{\Psi_\alpha}=\sum_j C_{j,\alpha}\ket{\rr_j}\,,
		\end{equation}
		can be associated to a given $\kk$-momentum of the Brillouin zone. To do it, one can perform the discrete Fourier transform of $C_{j,\alpha}$, i.e., $C_{\alpha(\kk)}=\sum_j C_{j,\alpha} e^{i\kk\cdot \rr_j}$, and see the $\kk$ that has the larger weight. 
		In this section, and in Fig. 1 of the main text, we use this approach to estimate the $\tilde{\bf{G}}(0)$  that needs to be included for a given finite size, and then renormalize with a small $a_{\mathrm{ho}}$ to get $\omega_\kk,\gamma_\kk$ is a more efficient way.
		
		\subsection{Results for in-plane polarized modes}
		
		\begin{figure}
			\centering
			\includegraphics[width=\linewidth]{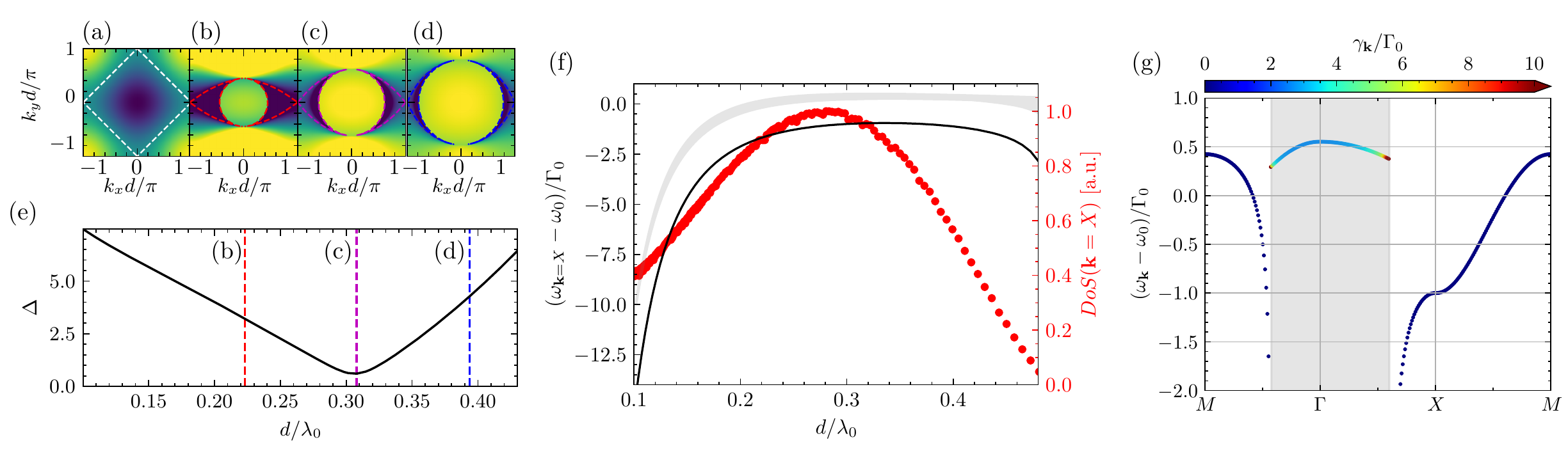}
			\caption{(a-d) Contour plot of $\omega_\mathbf{k}$ for (a) tight-binding model, and square atomic arrays with in-plane polarization (b) $d/\lambda_0=0.22$,  (c) $d/\lambda_0=0.31$ and (d) $d/\lambda_0=0.39$. The dashed lines are the iso-frequencies of the $X$-modes, i.e., $\omega_{\mathbf{k}=X}$. (e) Parameter $\Delta$ characterizing the proximity to the tight-binding model (see Eq.~\eqref{eqSM:Delta}). The vertical dashed lines denotes the distances for which we plotted the band-structure in panels (b-d). (f) Position of the Van Hove singularity (solid black line) and density of states at the $X$-mode energy (red circles) as a function of $d/\lambda_0$. The gray shadow region denotes the superradiant band. (g) Band structure for an in-plane polarized atomic dipole array with interatomic distance of $d/\lambda_0 = 0.3$. The colors represent the collective free space decay rate, $\gamma_\kk$, and the gray shadow region denotes the light cone.}
			\label{figSM:1-inplaneband}
		\end{figure}
		
		In the main text we choose $d/\lambda_0=0.3$ as the \emph{optimal} inter-atomic distance to obtain the directional behaviour for the in-plane polarized modes. The justify this choice we include here in Fig.~\ref{figSM:1-inplaneband} a systematic study of the metasurface band-structure $\omega_\kk$ for several distances. There, we first plot in Fig.~\ref{figSM:1-inplaneband}(a), to have it as a reference, the contour plot of the energy dispersion for the nearest-neighbour tight-binding model, i.e., $\omega_\kk/J=-2(\cos(k_x d)+\cos(k_y d))$, highlighting in dashed white the iso-frequency at the $X$-mode energy responsible of the directional emission~\cite{galve17a,Gonzalez-Tudela2017b,Gonzalez-Tudela2017a}. In ~\ref{figSM:1-inplaneband}(b-d) we plot the emergent band-structure of the metasurface for several distances, where we observe how the \emph{straightness} of the isofrequencies varies with the parameter $d/\lambda_0$. To characterize the straightness, we define a parameter that quantifies the difference of the isofrequencies with respect to the tight-binding model ones as follows:
		\begin{equation}
			\Delta = \int_0^{\pi/d} dk_x [(-k_x + \pi/d) - \tilde{k}_y(k_x)]^2\,,\label{eqSM:Delta}
		\end{equation}
		where $\tilde{k}_y(k_x)$ is the iso-frequency contour line obtained by solving $\omega(k_x, \tilde{k}_y) = \omega_{\mathbf{k}=X}$. In Fig.~\ref{figSM:1-inplaneband}(e) we can see that there is a lattice constant $d/\lambda_0\sim 0.31$ that minimize the distance to the tight-binding iso-frequencies, which remarkably, is similar to the one where the highest value of the density of states is, see Fig.~\ref{figSM:1-inplaneband}(f).

		\subsection{Results for other polarizations: out-of-plane and circularly polarized transitions}
		
		\begin{figure}
			\centering
			\includegraphics[width=1\linewidth]{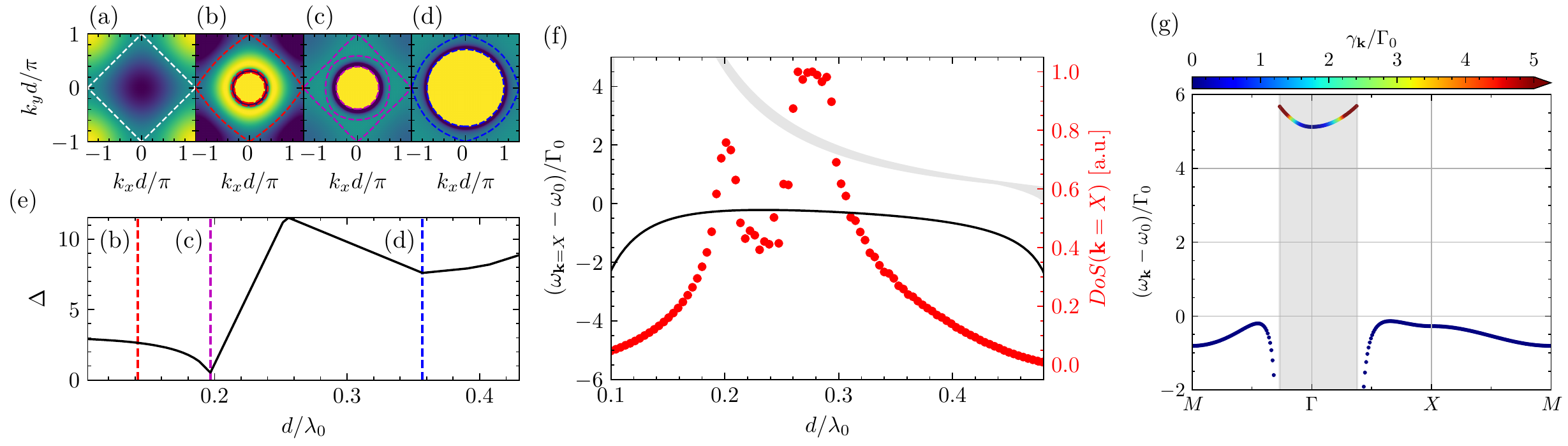}
			\caption{(a-d) Contour plot of $\omega_\mathbf{k}$ for (a) tight-binding model, and square atomic arrays with out-of-plane polarization $\hat{\boldsymbol{\wp}}_0 = \hat{e}_z$ for (b) $d/\lambda_0=0.1$,  (c) $d/\lambda_0=0.19$ and (d) $d/\lambda_0=0.36$. The dashed lines are the iso-frequencies of the $X$-modes, i.e., $\omega_{\mathbf{k}=X}$. (e) Parameter $\Delta$ characterizing the proximity to the tight-binding model (see Eq.~\eqref{eqSM:Delta}). The vertical dashed lines denotes the distances for which we plotted the band-structure in panels (b-d). (f) Position of the Van Hove singularity (solid black line) and density of states at the $X$-mode energy (red circles) as a function of $d/\lambda_0$. The gray shadow region denotes the superradiant band. (g) Band structure for an out-of-plane polarized atomic dipole array with interatomic distance of $d/\lambda_0 = 0.19$. The colors represent the collective free space decay rate, $\gamma_\kk$, and the gray shadow region denotes the light cone.
			}
			\label{figSM:outofplane}
		\end{figure}
		
		For completeness, we also include here a similar study to the one performed in Fig.~\ref{figSM:1-inplaneband}, but for the out-of-plane modes. This is what we show in Fig.~\ref{figSM:outofplane}(a-d), where we show the emergent band structure, $\omega_\kk$, of a subwavelength atomic array with $\hat{\boldsymbol{\wp}}_{0}=\hat{\bf{e}}_z$ for several distances $d/\lambda_0$. As in the case in-plane polarized situation, $\omega_\kk$ displays saddle points at the X-point and a straightening of their corresponding isofrequencies. This is more evident when studying the difference of these isofrequencies with respect to the nearest-neighbour model defined by the parameter $\Delta$ (see Eq.~\eqref{eqSM:Delta}), as shown in Fig.~\ref{figSM:outofplane}(e). There, we observe how the optimal distance is $d/\lambda_0\approx 0.2$ (see Fig.~\ref{figSM:outofplane}(g) for a line cut of $\omega_\kk$), also corresponds to a peak in the density of states, see Fig.~\ref{figSM:outofplane}(f). At larger $d/\lambda_0$ the model displays even a larger value of the density of states, however, this is also partly because other modes become resonant which will spoil directional emission, see Figs.~\ref{figSM:outofplane}(b-d). Although this larger density of states leads to larger impurity coupling factors (as we will see in the next section), we decided to keep the discussion in the main text for the in-plane polarized modes since they have a less demanding optimal distance ($d/\lambda_0\approx 0.3$) and a more tunable directionality.

		\begin{figure}
			\centering
			\includegraphics[width=\linewidth]{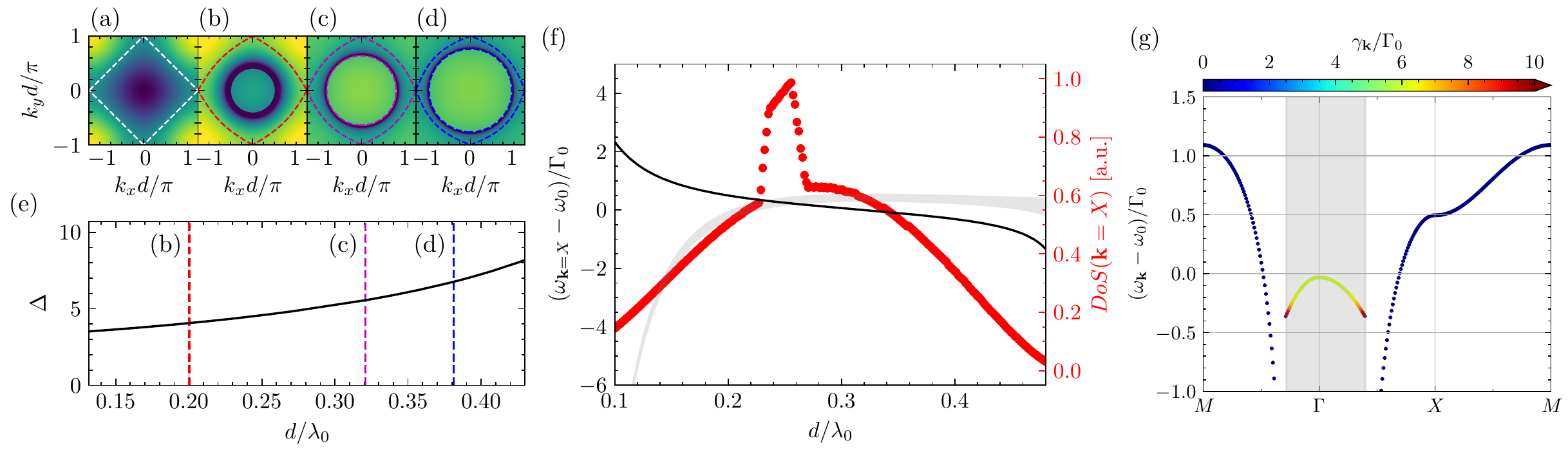}
			\caption{(a-d) Contour plot of $\omega_\mathbf{k}$ for (a) tight-binding model, and square atomic arrays with circularly polarized light polarization $\hat{\boldsymbol{\wp}}_0 = (\hat{e}_x + i\hat{e}_y)/\sqrt{2}$ for (b) $d/\lambda_0=0.2$,  (c) $d/\lambda_0=0.32$ and (d) $d/\lambda_0=0.38$. The dashed lines are the iso-frequencies of the $X$-modes, i.e., $\omega_{\mathbf{k}=X}$. (e) Parameter $\Delta$ characterizing the proximity to the tight-binding model (see Eq.~\eqref{eqSM:Delta}). The vertical dashed lines denotes the distances for which we plotted the band-structure in panels (b-d). (f) Position of the Van Hove singularity (solid black line) and density of states at the $X$-mode energy (red circles) as a function of $d/\lambda_0$. The gray shadow region denotes the superradiant band. (g) Band structure for an out-of-plane polarized atomic dipole array with interatomic distance of $d/\lambda_0 =0.2$. The colors represent the collective free space decay rate, $\gamma_\kk$, and the gray shadow region denotes the light cone.}
			\label{figSM:circular}
		\end{figure}
		
		Finally, in Fig.~\ref{figSM:circular} we include the same analysis for the case where the optical transition is circularly polarized: $\hat{\boldsymbol{\wp}}_{0}=(\hat{\bf{e}}_x+i\hat{\bf{e}}_y)/\sqrt{2}$. From the analysis of the band-structures and the $\Delta$-parameter, see Figs.~\ref{figSM:circular}(a-f), the iso-frequencies become straighter, the smaller $d/\lambda_0$, i.e., there seems not to be an optimal one. However, for large $d/\lambda_0$, the superradiant modes become resonant with the subradiant $X$-modes which can be detrimental for applications. In Fig.~\ref{figSM:circular}(g), we finally provide an example of band-structure for $d/\lambda_0=0.2$ showing how indeed it has a similar qualitative shape than the linearly-polarized modes.

		\section{Characterizing single impurity coupling to the metasurface~\label{secSM:impuritycoupling}}
		
		In this section, we will introduce an additional impurity atom together with the metasurface, as depicted in Fig. 1(a) of the main text. We denote as $\omega_a$ and $\boldsymbol{\wp}_{a}$ to the frequency and polarization vector of its optical transition, such that its individual free-space decay rate reads $\Gamma_a=|\boldsymbol{\wp}_a|^2\omega_a^3/(3\pi\hbar c^3)$. Thus, its intrinsic effective Hamiltonian can be written as $H_a/\hbar=(\omega_a-i\Gamma_a/2)\sigma^a_{ee}$. On top of that, this atom will exchange excitations with the metasurface through the same photon-mediated interactions appearing within the metasurface. The Hamiltonian describing such interaction is given by:
		\begin{align}
			\frac{H_{am}}{\hbar}=\sum_{j=1}^N\left(J_{aj}-i\frac{\Gamma_{aj}}{2}\right)\left(\sigma_{eg}^a\sigma_{ge}^j + \sigma_{eg}^j\sigma_{ge}^a\right)\,,
		\end{align}
		where $J_{aj},\Gamma_{aj}$:
		\begin{equation}
			J_{aj}-i\frac{\Gamma_{aj}}{2}=-\frac{3\pi \sqrt{\Gamma_0\Gamma_a}}{\omega_0}\hat{\boldsymbol{\wp}}^*_{a} \cdot\mathbf{G}_0(\rr_a-\rr_j)\cdot\hat{\boldsymbol{\wp}}_{j}\,.
			\label{eqSM:Green_tensorb}
		\end{equation}
		
		Note that here there is already an implicit assumptions that $|\omega_a-\omega_0|\ll\omega_a$ to replace all the occurrences of $\omega_a$ by $\omega_0$. 
		
		In what follows we are interested in characterizing the ratio of emission into the subradiant modes of the array ($\Gamma_m$) compared to the rest of the channels, e.g., free-space emission ($\Gamma'$). We denote this ratio as $P=\Gamma_m/\Gamma'$, which we label as the Purcell factor of the metasurface. In what follows we discuss two different methods to calculate it: a semi-analytical one in section~\ref{subsecSM:semianal} and a full numerical approach in section~\ref{subsecSM:numerical}. These are the methods used to calculate the results of Figs. 2-3 of the main text. Here, for completeness, we will also use it to characterize the coupling of an impurity with an out-of-plane polarized array in section~\ref{subsecSM:outofplanecoupling}.

		\subsection{Semi-analytical calculation of the impurity-metasurface coupling~\label{subsecSM:semianal}}
		
		Here, we extend the semi-analytical approach derived in Ref.~\cite{Masson2020} for 1D subwavelength arrays to the metasurface situation. There, it was found that assuming a Born-Markov approximation for the impurity atom dynamics, the emission into the metasurface and free space modes can be calculated as follows:
		\begin{align}
			\frac{\Gamma_m}{\Gamma_a}&=\frac{9d^2}{2k_a^2}\Im\left(\iint _{\abs{\mathbf{k}}> k_0}d^2\mathbf{k}\frac{\hat{\boldsymbol{\wp}}^*_a \boldsymbol{\alpha}_\mathbf{k}(\mathbf{r}_a)\otimes \boldsymbol{\beta}_\mathbf{k}(\mathbf{r}_a)\hat{\boldsymbol{\wp}}_a}{(\omega_a-\omega_\mathbf{k})/\Gamma_0}\right)\,,\label{eqSM:Gamma_m}\\
			\frac{\Gamma'}{\Gamma_a}&= 1 + \frac{9d^2}{2k_a^2}\Im\left(\iint _{\abs{\mathbf{k}}\leq k_0}d^2\mathbf{k}\frac{\hat{\boldsymbol{\wp}}^*_a \boldsymbol{\alpha}_\mathbf{k}(\mathbf{r}_a)\otimes \boldsymbol{\beta}_\mathbf{k}(\mathbf{r}_a)\hat{\boldsymbol{\wp}}_a}{(\omega_a-\omega_\mathbf{k})/\Gamma_0}\right)\,,
			\label{eqSM:Gamma_prime}
		\end{align}
		where $\mathbf{r}_a$ is the position of the impurity atom, which is where we evaluate the eigen-modes $\boldsymbol{\alpha}_\mathbf{k}(\mathbf{r}),\boldsymbol{\beta}_\mathbf{k}(\mathbf{r}) $ defined as:
		\begin{subequations}
			\begin{align}
				\boldsymbol{\alpha}_\mathbf{k}(\mathbf{r}) &= \sum_{j=1}^N\mathbf{G}_0(\mathbf{r}, \mathbf{r}_j, \omega_0)\cdot \hat{
					\boldsymbol{\wp}}_0e^{i\mathbf{r}_j\cdot\mathbf{k}},\\
				\boldsymbol{\beta}_\mathbf{k}(\mathbf{r}) &= \sum_{j=1}^N\hat{
					\boldsymbol{\wp}}_0^*\mathbf{G}_0(\mathbf{r}_j, \mathbf{r}, \omega_0)\cdot e^{-i\mathbf{r}_j\cdot\mathbf{k}}.
			\end{align}
		\end{subequations} 
		
		In Ref.~\cite{Masson2020} an analytical expression was found for both $\Gamma_m$ and $\Gamma'$ thanks to the 1D character of the modes. In our case, we can not find that expression, and we have to numerically calculate these integrals to obtain these values. We label as:
		\begin{align}
			P_a=\frac{\Gamma_m}{\Gamma'}\,,\label{eqSM:Q1anal}
		\end{align}
		to the single impurity-metasurface Purcell factor calculated using this semi-analytical approach. 
		
		\subsection{Numerical calculation of impurity-metasurface coupling~\label{subsecSM:numerical}}
		
		The other approach consists in initializing the impurity atom in the excited state, $\ket{\Psi(0)}=\sigma_{eg}^a\ket{\mathrm{vac}}$, and then compute the exact evolution of the system using the complete non-hermitian Hamiltonian of the combined impurity-metasurface system:
		\begin{align}
			\ket{\Psi(t)}=e^{-i H t/\hbar}\ket{\Psi(0)}\,,
		\end{align}
		where $H=H_a+H_m+H_{am}$. In this method we face two difficulties: i) how to avoid that the quantum emitter dynamics is affected by back-reflection of the subradiant excitations in finite systems; ii) how to extract the ratio $P$ from $\ket{\Psi(t)}$. 
		
		To tackle the first issue we impose non reflecting boundary conditions by including additional local losses in the dipoles close to the borders of the array. These additional losses must increase smoothly toward the ends of the lattices. For our calculations we find that a quadratic function is sufficient to avoid back scattering in most cases. The individual losses are given by:
		\begin{equation}
			\Gamma_0(\mathbf{r}_j) = \left\{\begin{array}{lr}
				\Gamma_0,& \quad \abs{\mathbf{r}_j} < r_\text{min}\\
				\Gamma_0 + \dfrac{\Gamma_\text{max}(\abs{\mathbf{r}_j} - r_\text{min})^2}{(r_\text{max} - r_\text{min})^2},& \quad \abs{\mathbf{r}_j}\geq r_\text{min}
			\end{array}\right.
		\end{equation}
		with $r_\text{min}$ the distance from the center of the lattice at which the individual losses starts to increase, and $r_\text{max}$ the maximum distance from the outermost dipole in the array. We empirically find that the best results in our system are obtained with $\Gamma_\text{max} = 10\Gamma_0$ and $r_\text{max} - r_\text{min} = 10d$. 
		
		In order to address the second issue, we use the fact that since $H$ conserves the number excitations, the wavefunction $\ket{\Psi(t)}$ at any time can be written as:
		\begin{align}
			\ket{\Psi(t)}=\left(C_a(t)\sigma^a_{eg}+C_{\rr_j}(t)\sigma^j_{eg}\right)\ket{\mathrm{vac}}\,,
		\end{align}
		from which we can monitor the total population in the impurity atom/lattice, i.e., $\mathcal{P}_a(t)=|C_a(t)|^2$ and $\mathcal{P}_L(t)=\sum_j|C_{\mathbf{r}_j}(t)|^2$. The effective Hamiltonian induces a non-unitary dynamics, since the excitations can be eventually lost in free space. Thus, although $\mathcal{P}_a(t=0)=1$, the sum $\mathcal{P}_a(t)+\mathcal{P}_L(t)<1$ for $t>0$, where the difference with the initial population $L_T(t)=1-\mathcal{P}_a(t)-\mathcal{P}_L(t)$ is the total loss into free-space of the system. This total loss can be also obtained by diagonalizing the quantum jump part of the master equation derived in Eq.~\ref{eqSM:mequation}, i.e., $\bf{\Gamma_{ij}}\ket{\phi_\alpha}=\gamma_\alpha \ket{\phi_\alpha}$, which leads to the following instantaneous quantum jump loss:
		\begin{equation}
			L_T(t) =\sum_\alpha L_\alpha (t)=\sum_{\alpha} \int _0^t dt'\gamma_\alpha \abs{\braket{\phi_\alpha}{\Psi(t')}}^2.
		\end{equation}
		
		If the metasurface modes were perfectly subradiant, then the Purcell factor could be obtained just by dividing $P(t)=\mathcal{P}_L(t)/L_T(t)$, since $L_T(t)$ will give the decay of the impurity atom to free-space. In the simulation, however, we introduce absorbing boundaries, that we include in $\Gamma_0(\mathbf{r}_j)$. These boundaries induce an additional decay of the metasurface modes to free-space that is artificially introduced by the simulation. Thus, it should be taken into account into the Purcell factor calculation to obtain a more accurate comparison with the semi-analytical procedure of Eqs.~\eqref{eqSM:Gamma_m}-\eqref{eqSM:Gamma_prime}, which assumes an infinite size system. To estimate how much of the total losses, $L_T(t)$, are lost via the adiabatic absorbing boundaries, we assume that the hybridization of the absorber atoms with the rest of the metasurface is weak, such that,  the losses in the border are given by
		\begin{align}
			L_b(t)\approx \sum_{\abs{\mathbf{r}_j} \geq r_{\textrm{min}}}\int_0^t dt' [\Gamma_0(\mathbf{r}_j) - \Gamma_0]\abs{C_{\mathbf{r}_j}(t')}^2\,.
		\end{align}
		
		With that, we can add/substract these "artificial" losses to each of the parts of the Purcell factor as follows:
		\begin{align}
			P(t)\approx \frac{\mathcal{P}_L(t) + L_b(t)}{L_T(t) - L_b(t)}\,.
		\end{align}
		
		When $t\rightarrow \infty$ this value converges to a finite value, which is what we define as the numerical Purcell factor:
		\begin{align}
			P_n=P(t\rightarrow\infty)\,.
		\end{align}
		
		This procedure is the one we use in the main text to compare with the semi-analytical approach of Eq.~\eqref{eqSM:Q1anal}. This method has several important advantages:
		\begin{itemize}
			\item It does not rely on Born-Markov approximation, and thus will be able to characterize impurity-metasurface couplings at all regimes.
			\item It can be easily generalized to calculate the Purcell factor of more general situations, such as the entangled cluster or bilayer configuration.
		\end{itemize}
		
		As we already observe for the in-plane polarized modes studied in the main text, both $P_a$ and $P_n$ display the same qualitative dependence with the system parameters, such as impurity-metasurface separation. The main quantitative difference appears in the regions where the Born-Markov approximation breaks, and the dynamics deviates significantly from a pure exponential decay. We explain this more in detail in the next section, where we characterize the impurity metasurface coupling for the out-of-plane transitions.
		
		\subsection{Results for out-of-plane emitters: non-Markovian dynamics~\label{subsecSM:outofplanecoupling}}
		
		\begin{figure}
			\centering
			\includegraphics[width=0.9\linewidth]{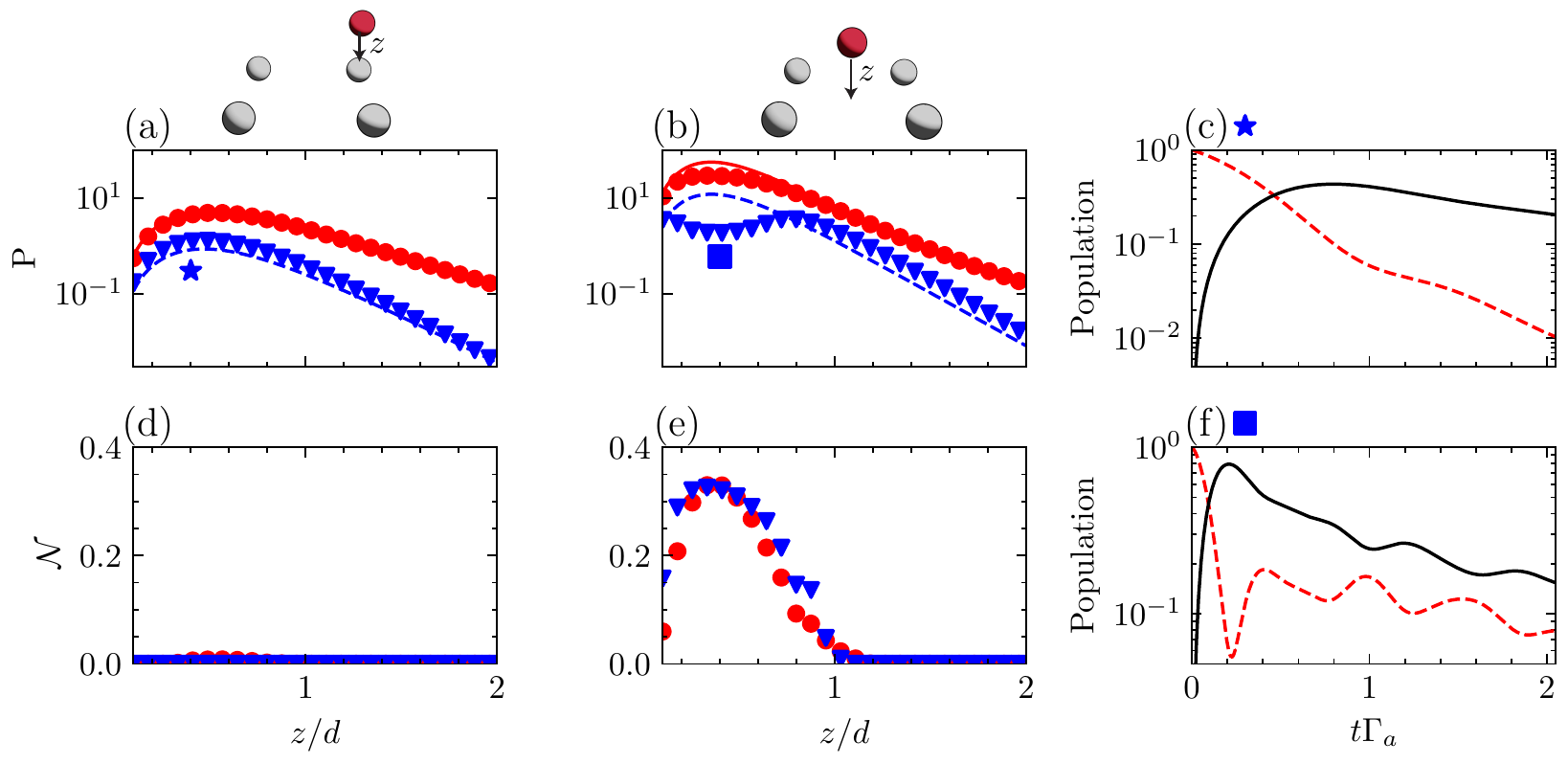}
			\caption{(a-b) Purcell factor $P$ computed semi-analytically (colored lines), and numerically (markers) for an impurity atom with $\hat{\boldsymbol{\wp}}_a = \hat{\mathbf{e}}_y$ placed nearby an array with $d/\lambda_0=0.19$ (solid red lines, red points), and $d/\lambda_0=0.28$ (dashed blue lines, blue triangles) with out-of-plane polarization $\hat{\boldsymbol{\wp}}_0 = \hat{\mathbf{e}}_z$. In (a) the emitter is located on top of a dipole, while in (b) is in the center of a plaquette. (d-e) Non-Markovian withness $\mathcal{N}$ defined in Eq.~\eqref{eqSM:NM} for the situations shown in the upper figures (a-b). (c) and (f) represent the excitation dynamics of the array (solid black line) and the impurity atom (dashed red line) with $z = 0.4d$ and a lattice constant of $d/\lambda_0=0.28$ for the two situations depicted with the markers in panels (a-b), corresponding to a typical Markovian/non-Markovian evolution, respectively.}
			\label{figSM:3-outofplanecoupling}
		\end{figure}
		
		In Fig.~\ref{figSM:3-outofplanecoupling}(a-b) we study the Purcell factor of an impurity atom with $\hat{\boldsymbol{\wp}}_a = \hat{\mathbf{e}}_y$ as a function of its vertical position for an array with out-of-plane polarized modes $\hat{\boldsymbol{\wp}}_0 = \hat{\mathbf{e}}_z$. In the two panels we plot the results for two different places of the unit cell depicted in the inset, as we did for the in-plane polarized modes in the main text. Compared to the case of the in-plane polarized modes, a noteworthy difference is that one can find regions with $P\approx 10$. This is expected since we already pointed that the narrower character of the band should lead to an increase of the density of states at the saddle-point divergence. Another important difference is that $P_a$ (dashed lines) and $P_n$ (markers) differ more than for the in-plane polarized metasurface. This points to the emergence of a more non-Markovian dynamics than in the in-plane polarized case. This is clear in Fig.~\ref{figSM:3-outofplanecoupling}(c) (Fig.~\ref{figSM:3-outofplanecoupling}(f)), we plot an example of the impurity atom dynamics in one of the points where the dynamics is (non-)Markovian, respectively. To make this more evident, in Figs.~\ref{figSM:3-outofplanecoupling}(d-e) we plot a non-Markovianity witness $\mathcal{N}$ of the dynamics introduced in Ref.~\cite{Lorenzo2017} for the same parameters of Fig.~\ref{figSM:3-outofplanecoupling}(a-b). This witness is defined as
		\begin{equation}
			\mathcal{N} = \frac{\int_{\partial_t |C_a(t)| > 0} dt \partial_t|C_a(t)|^4 }{\abs{\int_{\partial_t |C_a(t)| < 0} dt \partial_t|C_a(t)|^4}}\,,\label{eqSM:NM}
		\end{equation}
		in such a way that Markovian dynamics corresponds to $\mathcal{N} = 0$, and $\mathcal{N} = 1$ for a perfect coherent oscillations, the hall-mark of non-Markovian dynamics. Comparing panels (a-b) with (d-e) we observe how indeed the regions of larger deviations feature a larger value of the non-Markovian witness. Although not shown, this out-of-plane polarization also leads to cross-like directional patterns, though less tunable than the in-plane polarized modes. 
		
		
		
		\section{Directionality characterization~\label{secSM:tunabledirectional}}
		
		In order to characterize in more quantitative and qualitative terms the quasi-1D character of the emission patterns, we introduce a directional emission parameter, $\chi_{\mathrm{1D}}$. To define it, we will calculate the amount of emission passed by the dipoles located at a given distance from the center of the lattice $\abs{\mathbf{r}_i} \sim R$ where we place our impurity atom. Since our array is discretized, we pick those dipoles based on a midpoint circle algorithm~\cite{Agoston2005}. Then, we define the cumulative population of each of these selected dipoles as the integral over time
		\begin{equation}
			\mathcal{P}_j = \int_0 ^{\infty}dt |C_j(t)|^2.
		\end{equation}
		To compare the population of the dipoles in a concentric circle we redefine the cumulative population such that the sum over all dipoles in the circle is equal to one
		\begin{equation}
			\tilde{\mathcal{P}}_j = \dfrac{\mathcal{P}_j}{\sum_{\abs{\mathbf{r}_j}\sim R} \mathcal{P}_j}.
		\end{equation}
		
		Then we define the directionality as the sum of the populations for the dipoles located at a distance close to $R$, weighted by a function $W(\theta_j)$ which depends on the angle relative to the $x$-axis
		\begin{equation}
			\chi_{\mathrm{1D}} = \sum_{\abs{\mathbf{r}_j}\sim R} \tilde{\mathcal{P}}_j W(\theta_j).
		\end{equation}
		
		By working with the weight function $W(\theta_j) = \cos(2 (\theta_j - \theta_{\max}))$, where $\theta_{\max}$ is the angle of maximum population, we obtain that $\chi_{\mathrm{1D}}= 1$ if the emission is a straight line. In Fig. 2(c-d) of the main manuscript we fixed $\theta_{\max}$ always to $\pi/4$ for clarity of the figure, which is why we obtain negative values when the emission is orthogonal to it. Otherwise, if the emission is isotropic or cross-pattern like the nearest-neighbour hopping model, then $\chi_{\mathrm{1D}}= 0$. In order to mitigate the finite size effects, we compute $\chi_\mathrm{1D}$ for five concentric circles spaced at a distance of $d$, and obtain the mean value of the directionality.

		\section{Improvement strategies: entangled-clusters and bilayers~\label{secSM:improvement}}

		In this section, we provide more details on the strategies we sketch in the main text to achieve larger Purcell factors in 2D subwavelength arrays.

		\subsection{Entangled clusters~\label{subsecSM:improvement}}
		
		\begin{figure}
			\centering
			\includegraphics[width=\linewidth]{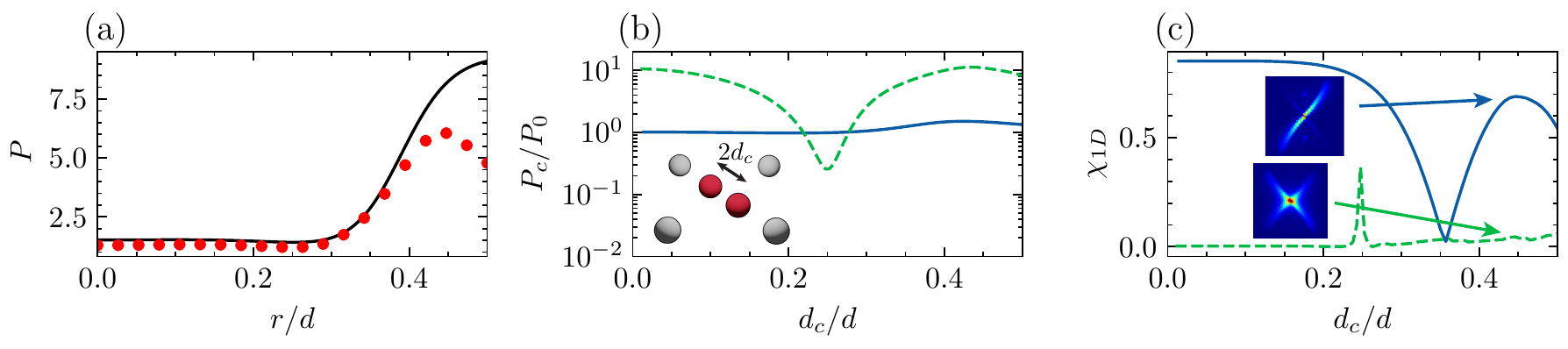}
			\caption{(a) Semi-analytical (solid back line) and numerical (red dots) Purcell factor of a single impurity as a function of the diagonal distance $r$ measured from the center of a plaquette. The impurity atom is located at $\mathbf{r}_a = (-r/\sqrt{2}, r/\sqrt{2}, 0)$. The inter-atomic distance of the array is $d/\lambda_0=0.3$, and the polarization $\hat{\boldsymbol{\wp}}_0 = \hat{\mathbf{e}}_y$ for lattice dipoles, and $\hat{\boldsymbol{\wp}}_a = (\hat{\mathbf{e}}_x + \hat{\mathbf{e}}_y) / \sqrt{2}$ for the impurity atom. (b) Numerical Purcell factor and (c) directionality for the entangled cluster configuration depicted in the inset of (b). Blue solid (dashed green) lines represent the cluster in the (anti-)symmetric state. Insets of (c) show the emission patterns in real space at the configurations denoted by the arrows. Horizontal and vertical axis represent the x and y directions, respectively. All other parameters are the same than in Fig. 3 (a) of the main text.}
			\label{figSM:5-entangledcluster}
		\end{figure}
		
		The first strategy is to consider two impurity atoms separated at a subwavelength scale instead of a single one. We denote the distance from each impurity atom to the center of the cluster with $d_c$, that will be subwavelength $2d_c<\lambda_a$. The intuitive idea is that if we prepare this atomic cluster in an entangled state: 
		\begin{align}
			\ket{\Psi_{c,\pm}(0)}=\frac{\ket{e_{a_1}g_{a_2}}\pm\ket{g_{a_1}e_{a_2}}}{\sqrt{2}}\,,
		\end{align}
		then the free-space emission between these two atoms can destructively interfere and boost the Purcell factor of the cluster, that we label as $P_c$. In Fig.~\ref{figSM:5-entangledcluster}(b-c) we calculate both $P_c$ and the associated directionality for the entangled cluster situation considered in the main text, and including both the symmetric (blue solid) and anti-symmetric configuration (dashed green) of the cluster. There, we observe how indeed the antisymmetric configuration can effectively achieved much larger $P_c$'s, but at expense of losing completely the tunability of the emission (though keeping the cross-directional emission pattern). We complement these figures with Fig~\ref{figSM:5-entangledcluster}(a) that shows the individual Purcell factor $P$ for each of the impurity atoms within the cluster as they move from the center of the plaquette. There, we observe how the impurity atoms couple more efficiently to the metasurface around $r\approx 0.45d$, which is where we observe an increase of $P$ for both the symmetric/antisymmetric configuration.

		\subsection{Bilayer configuration~\label{subsecSM:Bilayer}}
		
		\begin{figure}
			\centering
			\includegraphics[width=0.8\linewidth]{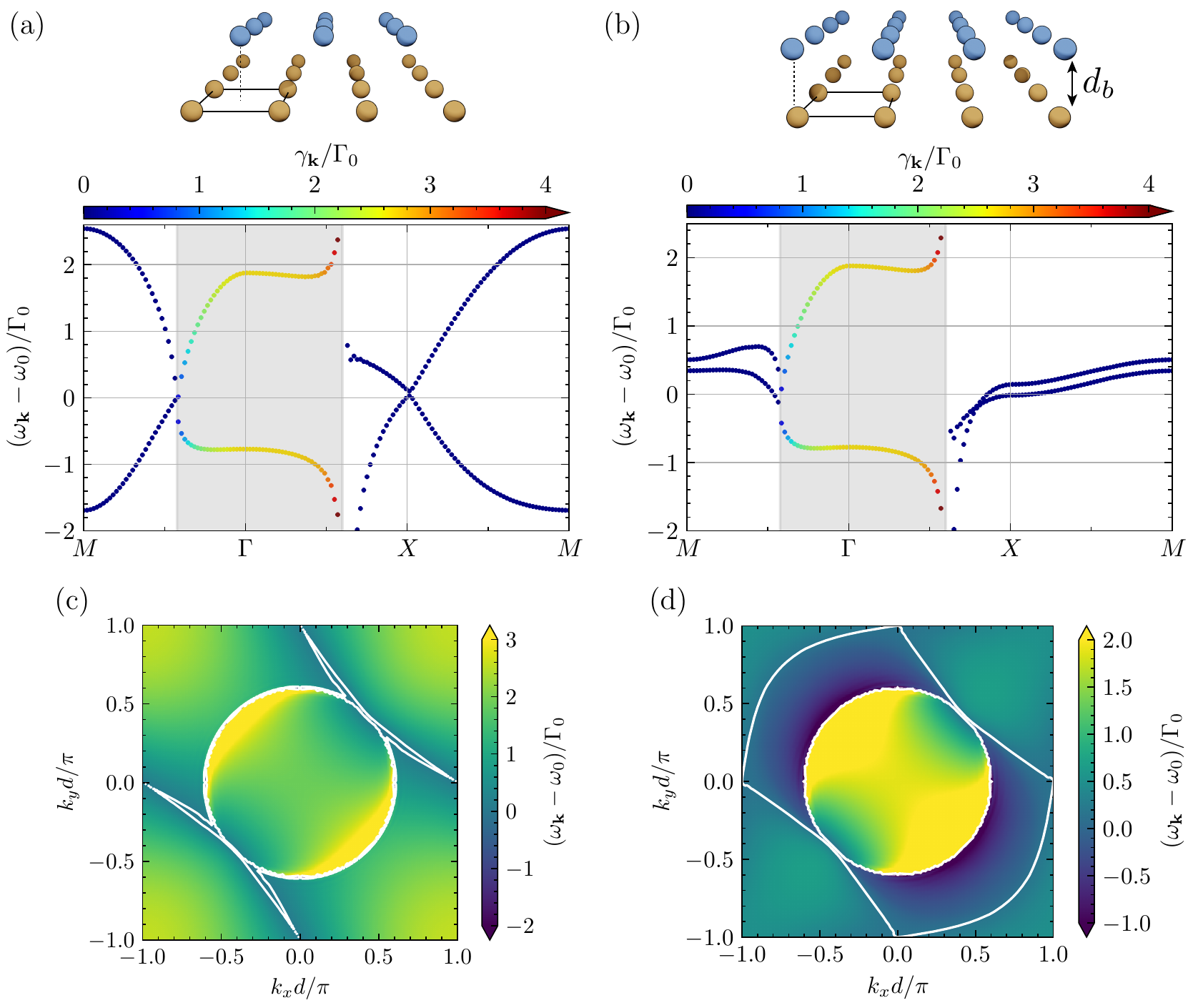}
			\caption{(a) Band structure for a bilayer with shift between layers of $\mathbf{d} = (0.5, 0.5)$, and a distance between them of $d_b = 0.1d$. (b) Band structure for a bilayer without shift between layers, and a distance between them of $d_b = d$. All the dipoles have a polarization parallel to the lattice $\hat{\boldsymbol{\wp}}_0 = (\hat{\mathbf{e}}_x + \hat{\mathbf{e}}_y) / \sqrt{2}$ and an interatomic distance of $d/\lambda_0 = 0.3$. In both cases, the colors represent the collective free space decay rate, and the gray shadow region denotes the light cone. (c-d) Contour plot of $\omega_\mathbf{k}$ for (a-b) respectively. The solid white line denotes the iso-frequencies of the $X$-mode.}
			\label{figSM:6-bilayerbandstructure}
		\end{figure}
		
		The other strategy considered is to change the single layer metasurface by a bilayer one. We assume that the two layers are displaced from the $z=0$ plane at distances $\pm d_b/2$, while we will keep the auxiliary atoms at the $z=0$ plane. Here, we characterize numerically the band-structure of the bilayer in Fig.~\ref{figSM:6-bilayerbandstructure}(a-b) considering two situations: one where the bilayers are displaced between them (a), and another one where the layers are top of each other (b). In both cases, one can find subradiant mode regions that around the $X$ points, one obtains straight iso-frequencies Fig.~\ref{figSM:6-bilayerbandstructure}(c-d) leading to directional emission patterns, as shown in Fig. 3(c) of the main text.
		
		\section{Collective dynamics~\label{secSM:collective}}
		
		In the main text, we explore the collective dynamics in three different initial configurations:
		\begin{itemize}
			\item The case of two separated single emitters at distance $d_e$ and initialized in a state:
			\begin{equation}
				\ket{\Psi_\mathrm{pair}(0)}=\frac{\ket{e_{a_1}g_{a_2}}+\ket{g_{a_1}e_{a_2}}}{\sqrt{2}}\,,
			\end{equation}
			\item The situation of two separated entangled clusters with intercluster separation $d_c$, but separated between them at the same position $d_i$:
			\begin{equation}
				\ket{\Psi_\mathrm{pair,c}(0)}=\frac{\ket{\Psi_{c_1,+}(0)}+\ket{\Psi_{c_2,+}(0)}}{\sqrt{2}}\,,\end{equation}
			
			\item The case of two separated atoms at distance $d_e$ among themselves and placed in-between a bilayer system with interlayer distance $d_b$:
			\begin{align}
				\ket{\Psi_\mathrm{pair,b}(0)}=\frac{\ket{e_{a_1}g_{a_2}}+\ket{g_{a_1}e_{a_2}}}{\sqrt{2}}\,,
			\end{align}
			
		\end{itemize}
		
		Using these initial states, we compute the dynamics using:
		\begin{align}
			\ket{\Psi_\alpha(t)}=e^{-i H t/\hbar}\ket{\Psi_\alpha(0)}\,,
		\end{align}
		where the index $\alpha$ denotes the different configurations. To estimate the strength of the collective effects we make a fit to an exponential decay law, i.e., $\ket{\Psi_\alpha(t)}\approx e^{-\Gamma_\alpha t}$, and plot $\Gamma_\alpha/\Gamma_\mathrm{ind}$, where $\Gamma_\mathrm{ind}$ is the value obtained by the individual decay at each configuration. In Fig.~\ref{figSM:7-dynamics}, we plot a representative example of such dynamics and their corresponding emission patterns.
		
		\begin{figure}
			\centering
			\includegraphics[width=0.5\linewidth]{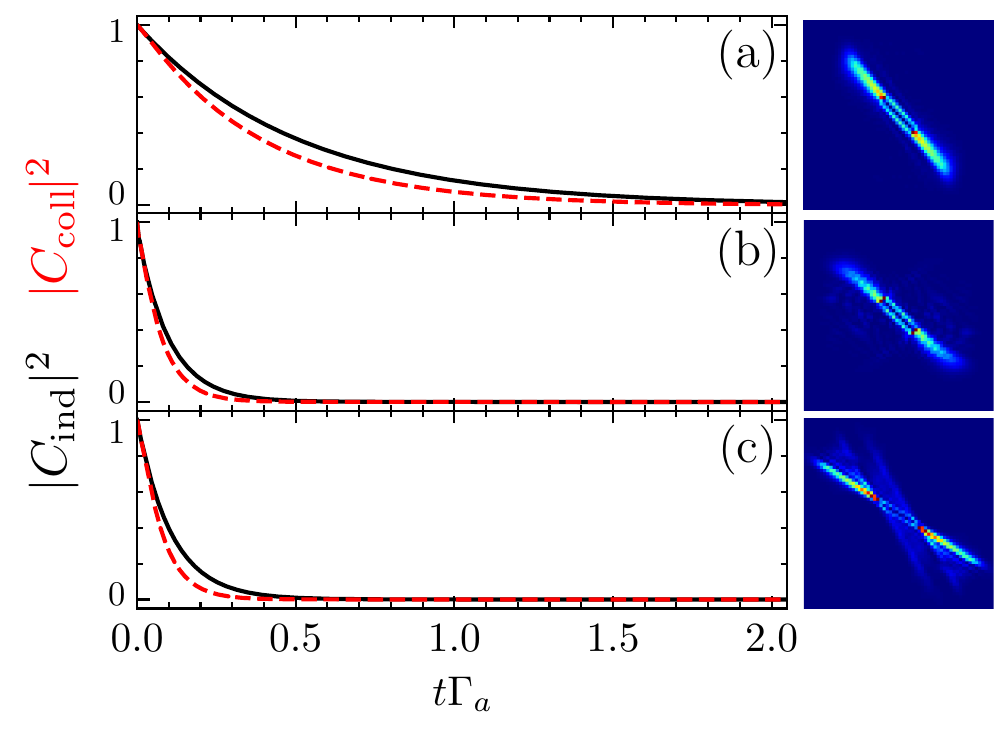}
			\caption{Individual (solid black lines) and collective (dashed red lines) time evolution for (a) single layer, (b) cluster, and (c) shifted bilayer. The emission pattern in real space for each case is plotted on the right, where the horizontal and vertical axis represent the $x$ and $y$ directions, respectively. Parameters: Lattice constant $d/\lambda_0 = 0.3$, cluster impurities distance $d_c=0.45d$, and bilayer distance $d_b = 0.1d$. The distance between emitters is $d_e = 14.14 d$ in (a-b), and $d_e = 15.26d$ in (c). Other parameters are the same than in Fig. 3 of the main text.}
			\label{figSM:7-dynamics}
		\end{figure}
		
		\section{Other geometries~\label{secSM:triangular}}
		
		The existence of Van Hove singularities and straight iso-frequency lines is ubiquitous in two-dimensional geometries. Thus, the shaping of the radiation patterns and collective decays that we predict along this manuscript can be further observed and tuned with other types of metasurfaces. As an example, we plot in Figs.~\ref{figSM:8-triangular-z} and  Figs.~\ref{figSM:8-triangular-x} the results obtained for a triangular geometry metasurface with interatomic spacing with perpendicular ($\hat{\boldsymbol{\wp}}_0 = \hat{z}$) and parallel ($\hat{\boldsymbol{\wp}}_0 = \hat{x}$) polarization vectors, respectively, see Fig.~\ref{figSM:8-triangular-z}(a) for a scheme. In Fig.~\ref{figSM:8-triangular-z}(c), we observe how the perpendicular polarized case indeed depicts straight iso-frequencies coinciding with the M-point, which is where the saddle point appear. The main difference with respect to the square metasurface is that the isofrequencies show an hexagonal shape instead of a square one. In Fig.~\ref{figSM:8-triangular-z}(b), we show how this point features a Van Hove singularity in the density of states, which leads to directional emission when an emitter is coupled resonantly to such energies, as depicted in Figs.~\ref{figSM:8-triangular-z}(d-e). The difference is that now that the emission occurs in six directions, instead of four, due to the different symmetry properties of the underlying geometry. In Fig.~\ref{figSM:8-triangular-z}(f-g), we further show how such emission can also be changed by changing the relative polarization between the impurity and the metasurface atoms, allowing to cancel partly the emission in some of the directions.

		\begin{figure}
			\centering
			\includegraphics[width=\linewidth]{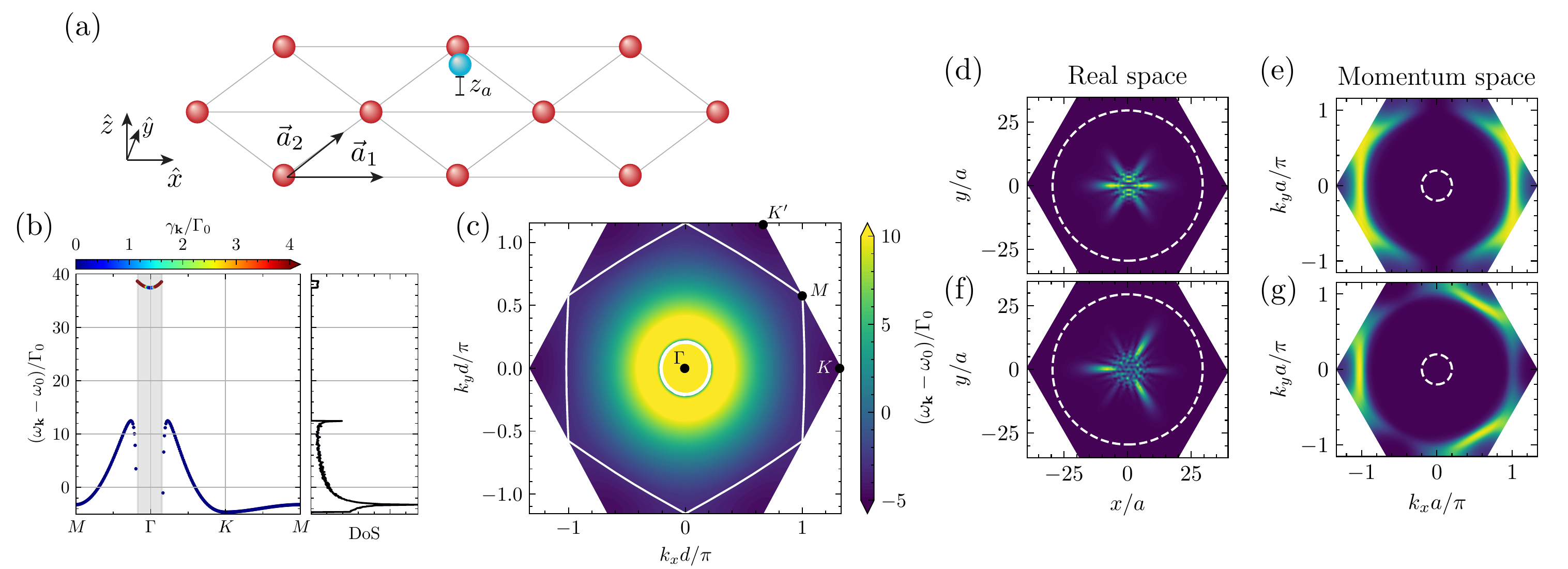}
			\caption{(a) Schematic picture of a triangular lattice (red spheres) with vectors $\vec{a}_1 = (d,0)$, $\vec{a}_2 = (d / 2, \sqrt{3} d/ 2)$, and an emitter (blue sphere) located in the center of a plaquette at height $z_a$. (b) Band structure along the principal symmetry points for a triangular lattice with interatomic spacing $d/\lambda_0 = 0.1$ and an out-of-plane polarization $\hat{\boldsymbol{\wp}}_0 = \hat{z}$. The color represent the free space decay rate and the light gray shadow region delimits the light cone. (c) Contour plot of $\omega_\mathbf{k}$ in the first Brillouin zone. The solid white line denotes the iso-frequencies of the M-mode. The right most figures show the emission patters in real space (d, f), and momentum space space (e, g) for an emitter resonant with the M-mode, with a polarization $\hat{\boldsymbol{\wp}}_a = \hat{x}$ (top row), and $\hat{\boldsymbol{\wp}}_a = \hat{x} + i\hat{y}$ (low row). The emitter is located at a height of $z_a=0.4d$ and its free space decay rate is $\Gamma_a = 0.02\Gamma_0$.}
			\label{figSM:8-triangular-z}
		\end{figure}
		
		For completeness, we also show how the emission and interactions can be further tuned by changing the polarization of the metasurface atoms to be parallel to the metasurface plane, i.e.,  $\hat{\boldsymbol{\wp}}_0 = \hat{x}$. This choice leads to a privileged direction and changes significantly the band-structure isofrequencies and density of states, see Fig. ~\ref{figSM:8-triangular-x}(a-b). In particular, one does not have an hexagonal shape for the M-point isoequifrequencies. However, we find quite straight one-dimensional lines in the isofrequencies of the $K'$ point, which enable quasi-1D and chiral emission depending on the relative orientation of the impurity atom polarization (see Figs.~\ref{figSM:8-triangular-x}(c-f)).
		
		\begin{figure}
			\centering
			\includegraphics[width=\linewidth]{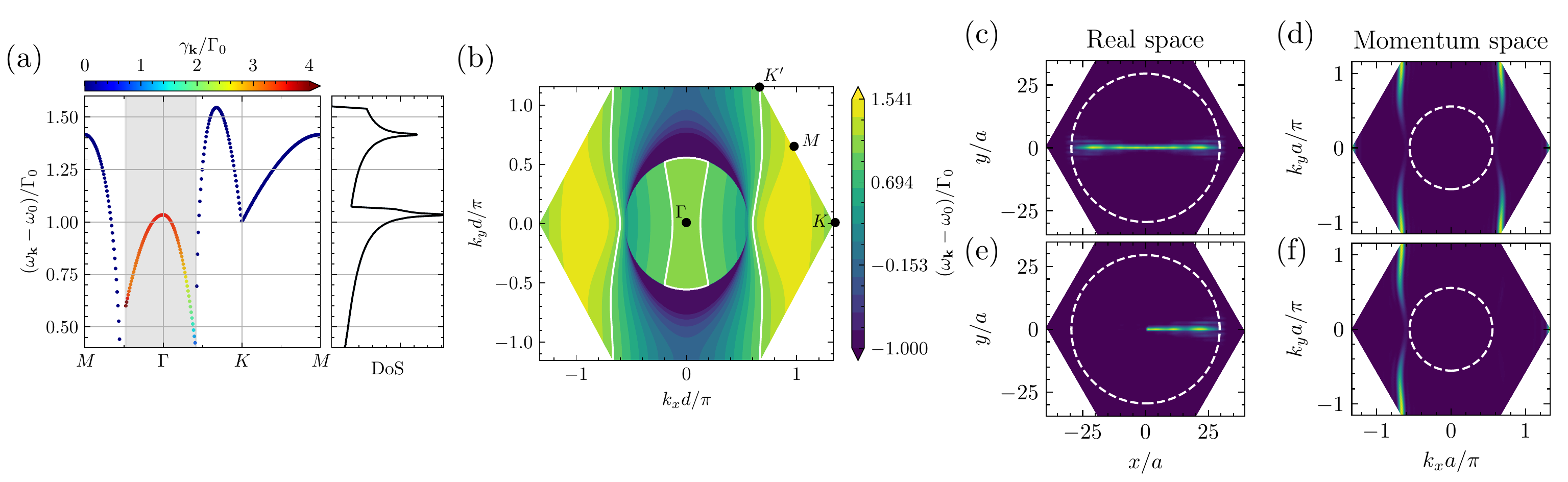}
			\caption{(a) Band structure along the principal symmetry points for a triangular lattice with interatomic spacing $d/\lambda_0 = 0.278$ and an in-plane polarization $\hat{\boldsymbol{\wp}}_0 = \hat{x}$. The color represent the free space decay rate and the light gray shadow region delimits the light cone. (b) Contour plot of $\omega_\mathbf{k}$ in the first Brillouin zone. The solid white line denotes the iso-frequencies of the K'-mode. The right most figures show the emission patters in real space (c, e), ans momentum space space (d, f) for an emitter located in the middle of a plaquette resonant with the K'-mode, with a polarization $\hat{\boldsymbol{\wp}}_a = \hat{x}$ (top row), and $\hat{\boldsymbol{\wp}}_a = \hat{x} + i\hat{y}$ (low row). The emitter is located in the plane of the lattice, and its free space decay rate is $\Gamma_a = 0.02\Gamma_0$.}
			\label{figSM:8-triangular-x}
		\end{figure}

		\section{Experimental considerations~\label{secSM:experimental}}
		
		Although there are already good references in the literature where they explain how to implement the setup explored along this manuscript~\cite{Masson2020,Patti2021,Brechtelsbauer2020,Castells-Graells2021a}, for completeness we will just review here how to obtain the different experimental ingredients required for it:
		
		\begin{itemize}
			\item \emph{Subwavelength arrays.} The key idea consists in using two different optical wavelengths for generating the optical lattice potentials that trap the atoms and for probing them. This was done, for example, in the recent experiment with Rb atoms~\cite{Rui2020} where they use an isolated two-level transition of the D2 line for probing ($\lambda_0=782$~nm) and another one for generating the optical lattice $\lambda=1064$~nm, being able to achieve $d/\lambda_0\approx 0.68$ since the distance between the minima in the optical lattice is $d=\lambda/2$. To be able to reach the deeply subwavelength regime, Alkaline-Earth atoms, such as Ytterbium and Strontium, are more promising since they combine telecom with near ultra-violet transitions in their atomic energy spectrum~\cite{ludlow15a}. For example, trapping Strontium at one its magic wavelengths~\cite{Olmos2013} and driving through the telecom transition $^3P_0\rightarrow ^3 D_0$ would enable $d/\lambda_0\approx 1/16$~\cite{Masson2020}, whereas using the Ytterbium transitions $^3P_0\rightarrow ^3 D_1$ ($\lambda=1.2\mu$m) and $^1S_0\rightarrow ^3 D_j$ ($\lambda\sim 470$\,nm)~\cite{covey19a}, one could obtain $d/\lambda\approx 0.2$.
			
			\item \emph{Impurity atoms control.} For the trapping of the additional auxiliary atoms, one can initially trap them far from the metasurface with optical tweezers~\cite{Kaufman2015}, and then move the traps near it as in recent realizations with nanophotonic structures~\cite{Dordevic2021}. Regarding the control of its properties (frequency and linewidth), one can use a Raman-assisted transition with a $\Lambda$-scheme~\cite{porras08a,Brechtelsbauer2020,Castells-Graells2021a}, containing two hyperfine ground state levels, labelled as $e,g$, connected to a common optically excited state $f$ (see Fig.~\ref{figSM:Lambda-scheme}). The connection between the $f\leftrightarrow e$ has to be done directly through a (detuned) classical laser field, or through a two-photon process, with overall amplitude $\Omega$ and detuning $\Delta=\omega_f-\omega_e-\omega_L+\omega_g$. The $f-g$ transition has to be a cycling transition in such a way that the free-space photons only induce decay through that channel. Then, under the conditions $|\Omega|,\Gamma_0 \ll |\Delta| $, the dynamics of the optically excited state can be adiabatically eliminated, arriving to an effective two-level transition between the $g\leftarrow e$ states which emits photons at frequencies $\sim \omega_L-(\omega_e-\omega_g)$, and with a renormalized linewidth $\sim \frac{|\Omega|^2}{\Delta^2}\Gamma_0$. Thus, the dynamics can be made more or less Markovian by tuning the Raman factor $|\Omega|^2/\Delta^2$. Relaxing the dynamical control requirement, another alternative is the use of other atomic isotopes which show similar atomic transitions but different linewidths~\cite{Masson2020,ludlow15a}.
		\end{itemize}
		
		\begin{figure}[tb]
			\centering
			\includegraphics[width=0.5\linewidth]{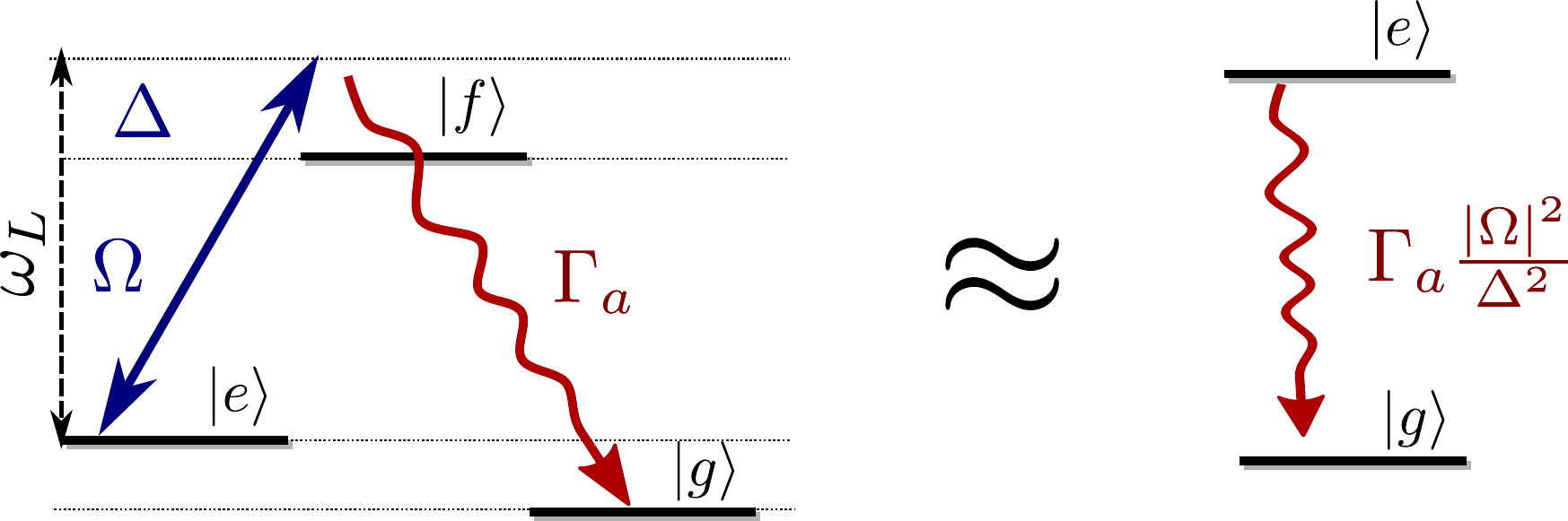}
			\caption{Sketch of a Lambda-Scheme with Raman-assisted transitions to achieve dynamical control of the frequency and linewidth of the impurity atoms. Two hyperfine ground state levels, $g$ and $e$, are connected via an optically excited state $f$. The coupling of the $e-f$ leg is done through a detuned Raman laser with amplitude $\Omega$ and detuning $\Delta$, whereas the $f-g$ leg is connected via a dipole transition through the free-space photonic bath. In the conditions where one can adiabatically eliminate the excited photonic state, $f$, i.e., $|\Omega|,\Gamma_0\ll \Delta$, the system is equivalent to a two-level optical transition $e-g$ with renormalized frequency $\omega_L-\omega_e+\omega_g$ and linewidth $\frac{|\Omega|^2}{\Delta^2}$.
				\label{figSM:Lambda-scheme}}
		\end{figure}
	\end{widetext}
	
	\bibliographystyle{apsrev4-2}
	\bibliography{referencesv2}
	
\end{document}